# Thin Ice Lithospheres and High Heat Flows on Europa From Large Impact Structure Ring-graben

**K. N. Singer[1], W. B. McKinnon[2], and P. M. Schenk[3]**

[1]Southwest Research Institute, Boulder, CO, USA. [2]Department of Earth, Environmental, and Planetary Sciences and McDonnell Center for Space Sciences, Washington University in St. Louis, Saint Louis, MO, USA. [3]Lunar and Planetary Institute, Houston, TX, USA.

Corresponding author: Kelsi Singer (kelsi.singer@swri.org)

**Key Points:**

- Circumferential ring-graben depths, widths, and spacing measured for the two largest known impact structures on Europa: Tyre and Callanish
- Graben widths give plausible depths to the brittle-ductile transition of ~2–3 km at the time of and under the conditions of impact
- Heat flows are high, dependent on fault structure and thermal conductivity, but consistent with impact breaching of a thin ice shell or lithosphere

**Abstract**

Craters are probes of planetary surface and interior properties. Here we measure depths, widths, and spacing of circumferential ring-graben surrounding the two largest multiring impact structures on Europa, Tyre and Callanish. We estimate formation conditions including the ice shell structure. The radial extension necessary to form these graben is thought to be caused by asthenospheric drag of warmer, more ductile ice and/or water flowing toward the excavated center of the crater, under a brittle-elastic lithospheric lid. Measurements of graben depths from stereo-photoclinometric digital elevation models result in estimates of displacement, strain, and stress experienced by the ice shell. Graben widths are used to estimate the intersection depth of the bounding normal faults, a quantity related to the brittle-ductile transition depth that approximates elastic shell thickness during crater collapse. Heat flows at the time of crater formation as well as ice lithosphere and total shell thickness are thus also constrained. Average widths and depths tend to decrease with increasing distance from the structure center, while inter-graben spacing generally increases. Varied assumptions yield plausible total conductive ice shell thickness estimates between 4–8 and 2.5–5 km for Tyre and Callanish, respectively, and heat flows of ~70–115 (±30) mW m$^{-2}$ for realistic thermal conductivities, consistent with other geophysical estimates for Europa. Higher heat flows are consistent with thin (≲10 km), conductive ice shells and impact breaching, or penetration of the stagnant lid for a convecting ice shell. Callanish, geologically younger, formed in a time or region of greater heat flow than Tyre.

**Plain Language Summary**

Jupiter's moon Europa has an outer icy shell overlying a global subsurface ocean. The thickness of this icy shell controls the appearance of impact craters. The shell thickness is not well known, and may change over time, but geological indicators suggest it is on the order of 10 km thick or more. This is a relatively thin layer for craters to form in, compared the hundred-km or thicker ice shells on most other icy bodies in the solar system. The two largest europan impacts created sets of circumferential ring-faults that take the form of graben/troughs in the outer portions of their structures. We use measurements of the graben widths and depths to derive minimum estimates of the ice lithosphere (upper brittle portion) and total shell thicknesses and heat flows on Europa at the time the structures formed. The measurements are consistent with a minimum ice shell thickness between 2.5 and 8 km and high to exceptionally high heat flows. Greater total ice shell thicknesses are permitted if the shell is convectively overturning at depth. Notably, our heat flow estimates for Callanish (the younger structure) are uniformly higher than for Tyre (the older), contrary to expectations for Europa.

**1. Introduction**

Clues to the ice shell thickness of Europa can be found in all of the geologic features that decorate its surface. Although impact craters are not the dominant feature shaping the young surface of Europa, the craters that are observed can place strong constraints on the ice shell thickness (e.g., Moore et al., 1998; Turtle, 1998; Moore et al., 2001; Turtle & Pierazzo, 2001; Schenk, 2002; Schenk & Turtle, 2009; Bray et al., 2014; Cox & Bauer, 2015; Silber & Johnson, 2017; Silber & Johnson, 2018). These constraints are possible because impact cratering as a geologic process is generally well understood from analogues across the solar system, and also because many aspects of the final crater morphology can be related to ice shell thickness at the time of the impact.

The largest impact features on Europa, Tyre and Callanish, do not have a classical rim crest or central peak, but instead exhibit multiple concentric ridges and graben-like troughs extending up to ~100 km away from their centers. Note that we will sometimes refer to these structures as basins herein, consistent with preexisting impact literature, but neither Tyre nor Callanish are broad, large-scale topographic lows (Schenk, 2002). Europan impact structure ring-graben are more compact but similar to ring-graben around the Valhalla and Asgard/Utgard multiringed impact structures on Callisto (McKinnon & Melosh, 1980; Schenk et al., 2004a), and plausibly related to the hemispherical-scale furrow systems on Ganymede (McKinnon & Parmentier, 1986; Schenk & McKinnon, 1987; Hirata et al., 2020). A relatively thinner ice shell on Europa presumably leads to these unique crater morphologies, and the onset of ring-graben at smaller impact sizes than on Callisto or Ganymede, given that the gravities of all 3 moons are similar.

Extension of the ice lithosphere is needed to produce ring-graben via brittle failure and normal fault offset (see general discussion of fault mechanics in Watters & Schultz, 2009; Melosh, 2011). This extension is thought to result from the lithosphere being dragged inward by a more mobile underlying asthenosphere (either warm, ductile ice, or water; **Figure 1**) that flows toward the crater center due to the pressure gradient created by the transient crater cavity as it collapses (Melosh & McKinnon, 1978; McKinnon & Melosh, 1980). Most of these processes are thought to occur relatively promptly after the impact in order to generate enough stress to create the graben. Transient cavity collapse is likely the most appropriate timescale and occurs on the order of several minutes for Tyre and Callanish (we return to this topic below).

The widths and depths of impact structure ring-graben can be used to derive the brittle ice shell thickness and heat flow at the time of their formation. The total ice shell thickness, the combination of both the upper brittle and lower ductile portions, can be estimated using those two derived quantities and information and/or constraints on the timescales and strain rates of the processes.

In this paper we examine images and topographic data for Tyre and Callanish, the two largest known impact structures on Europa. The "effective final diameter" ($D_{f,eff}$) of these two craters has been estimated by scaling from the ejecta deposits and onset distance of secondary craters (from Table 1 in Schenk & Turtle [2009], based on Schenk & Ridolfi [2002]). Tyre is slightly larger, estimated to be equivalent to a 38-km-diameter final crater (if formed on an otherwise similarly icy target, such as Ganymede) and Callanish is estimated to be closer to 33 km in effective diameter. The effective rim locations on both structures are inside of the first of what are prominent concentric tilted blocks/ridges, which themselves lie inward of the ring-graben or troughs. The transient crater diameter ($D_{tr}$), which is necessarily smaller than $D_{f,eff}$, is also a quantity of interest for understanding graben formation. Using the scaling in McKinnon & Schenk (1995) for complex craters on large icy satellites, specifically Ganymede but applicable to Europa given their similar crustal compositions and surface gravities, $D_{tr}$ = 23 and 20 km for Tyre and Callanish, respectively (cf. Cox & Bauer, 2015).

For completeness we note that the original *Galileo*-based morphological analysis of these two impact features (Moore et al., 2001) estimated "final" rim diameters directly from the limits of mapped inner smooth and rough units and/or their truncation or burial of preexisting tectonic structures (i.e., no actual rims or rim remnants are seen). Diameter estimates for Callanish ranged between 29 and 47 km, and that for Tyre was ≈44 km (Moore et al., 2001), which would (Tyre) or could (Callanish) imply larger transient craters than in **Table 1**. To the degree that this matters in our analysis and interpretations below, which is only slightly, these alternative sizes should be considered.

The structure of the paper is as follows. In Section 2 we detail our measurement methods, both the image and topographic data (Section 2.1) and the techniques used to measure them (Section 2.2). Results follow directly, on ring-graben widths (Section 3.1), depths (Section 3.2) and spacing (Section 3.3). An extensive Interpretation and Analysis section (Section 4) then attempts a deep look at what can be learned about the formation of Tyre and Callanish, and of the state of Europa' ice shell at the time of their impacts. Some aspects are relatively straightforward, in the sense that they rely mostly on surface observations, such as assessing the amount of inward radial displacement and strain each impact represents (Section 4.1), general interpretations of ring structure (Section 4.2), and timing constraints from ejecta and graben superposition relationships (Section 4.3). Results of prior numerical models are summarized in Section 4.4, followed by a re-look at the mechanics of ring-graben formation during a large impact (Section 4.5). This is all then used to constrain, as best as possible, the subsurface brittle-ductile transition and heat flow at the time and location of the Tyre and Callanish impacts (Section 4.6). These in turn constrain the total ice shell or lithospheric thicknesses at those times and places (Section 4.7). We conclude by highlighting the commonalities and differences between Tyre and Callanish (Section 5), and then offer a final summary and overview (Section 6).

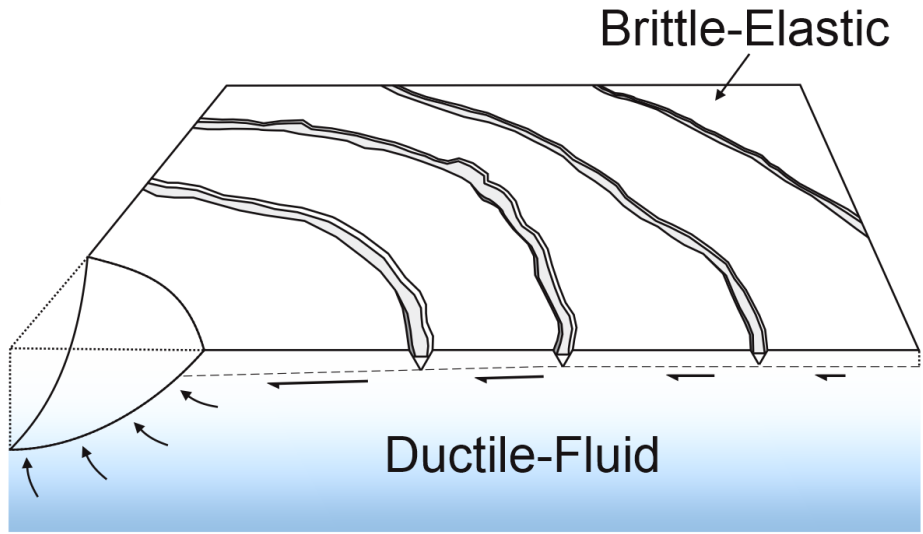

**Figure 1. Ring tectonic theory.** Illustration of the processes that may occur in the ice lithosphere and the warmer underlying asthenosphere or ocean directly after a large impact on Europa (following McKinnon & Melosh, 1980). Flow of the underlying more mobile material toward the cavity created by the impact may cause extensional faulting in the upper colder brittle part of the ice shell. Dashed line is the brittle-ductile transition, which deepens toward the cavity due to higher strain rates.

## 2. Measurement Methods

### 2.1. Image and Topographic Data

Tyre "Macula" was first observed by *Voyager 2*, with a best view of ~2 km px$^{-1}$, and was described as a likely degraded, ringed impact structure (technically, a crater palimpsest), even in lower resolution images; Callanish, however, was simply noted in *Voyager 1* imagery as a dark circular spot of similar scale and possibly similar nature (Lucchitta & Soderblom, 1982). The *Galileo* spacecraft Solid State Imager (SSI) obtained targeted images of Tyre and the southern half of Callanish (**Figures 2-3**). The Tyre mosaic is 170 m px$^{-1}$. A small, higher-resolution inset of ~30 m px$^{-1}$ is available to the southeast of Tyre but lies mostly outside of the graben region (**Figure 4d**). The larger Callanish mosaic utilized here is at a pixel scale of 120 m px$^{-1}$, whereas a smaller, mostly overlapping region is covered at 45 m px$^{-1}$. These two mosaics cover only the central and southern portions of Callanish. The best resolution image covering all of Callanish was taken by *Galileo* with Callanish near the terminator at the time of imaging at ~955 m px$^{-1}$ (**Figure 4b**). The higher resolution Callanish mosaic extends a bit farther north than the lower resolution mosaic and covers more area with prominent graben (example profile in **Figure 5**). The higher resolution mosaic and its associated topography were used for most of the measurements of the Callanish graben to the east.

We used topographic products produced primarily through the technique of albedo-controlled photoclinometry (see methods and discussion in Schenk, 2002; Schenk et al., 2004b; Schenk & Turtle, 2009; Singer et al., 2021) with some stereo control where possible. Photoclinometry relates the brightness of a pixel to the slope of the average surface in that pixel through a photometric function using the known lighting geometry. The intrinsic albedo of a surface can be somewhat controlled for by using higher-sun, lower-resolution images available for some areas of Europa. The Tyre digital elevation model (DEM) is derived from photoclinometry with some albedo control. Because photoclinometry estimates the slopes of the pixels and integrates across the scene to build up the DEM, the vertical precision is a function of distance and scale (Schenk, 2002; Schenk et al., 2004b; Bland et al., 2007; Singer et al., 2021). The accumulated slope-derived errors are not large for narrow features like the graben measured here. The vertical error per pixel in the topographic data set can be derived from the photoclinometric uncertainty in the slopes, which are ~1-to-2 degrees per pixel for the Tyre mosaic. This results in a vertical, or pixel height error of ~3-to-6 m per pixel at the resolution of the Tyre mosaic (~170 m px$^{-1}$). The error for a specific graben depth measured here is more difficult to quantify as each graben is unique, and small effects from both the error inherent in the photometric function or small-scale albedo variations may be present. If the surface is assumed to have fairly uniform photometric properties, random accumulating errors should be related to the pixel scale ($n$) as approximately $\sqrt{n}$ × the pixel height error. The Callanish DEM is a stereo-controlled photoclinometric product also with some albedo control. The vertical precision of the stereo component of the Callanish DEM is estimated through standard stereo techniques (Kirk et al., 2003) to be ~40 ± 10 m. As at Tyre, the photoclinometric component of the DEM improves the vertical precision of the DEM by roughly an order of magnitude (to approximately a few meters). The photoclinometry software interpolates in shadows to determine topography and the symmetry of most graben in the mapping site indicates that the height estimates are reliable to ~10% or better.

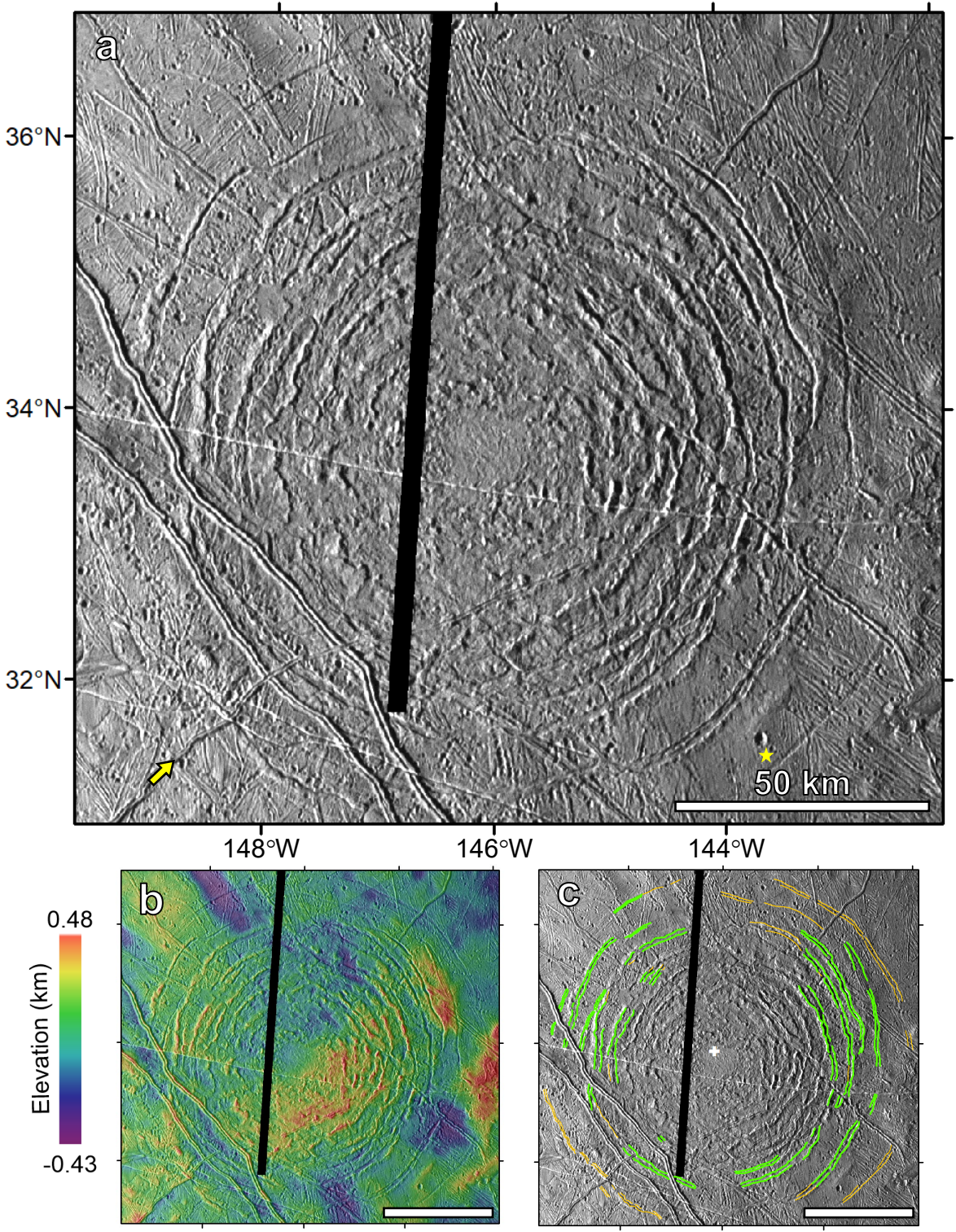

**Figure 2. Tyre multiring basin.** (a) Base mosaic (120 m px$^{-1}$). Arrow indicates a double ridge that transitions to a more of a fracture morphology as it crosses Tyre. Yellow star indicates location of higher resolution image shown in Figure 4d. (b) Color-coded topography. (c) Graben as mapped for this study. Thicker green outlines indicate more prominent graben walls, while thinner gold outlines indicate more subtle features or features with only one apparent scarp. White plus

sign is the fitted center of the basin (see text for fitting methods), which appears offset from the smoother-textured central unit. The geographic (lat-long) extent is the same in all of the panels and scale bar is 50 km in all cases.

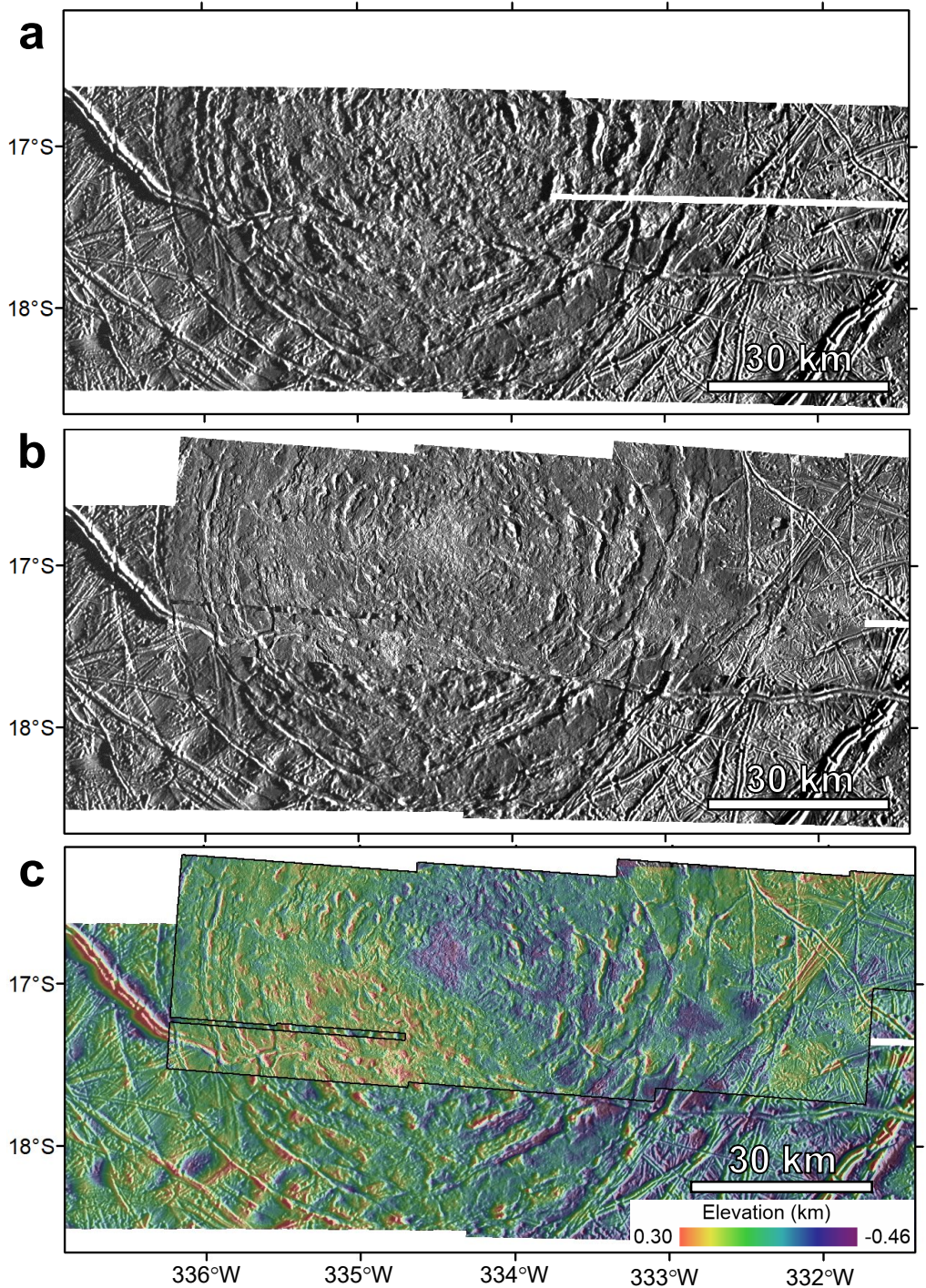

**Figure 3. Callanish multiring basin.** (a) Lower-resolution base mosaic (120 m px$^{-1}$). (b) Higher-resolution mosaic (45 m px$^{-1}$) overlain on panel (a). (c) Color-coded topography overlain on both mosaics. Mapped graben can be seen in Figure 6b.

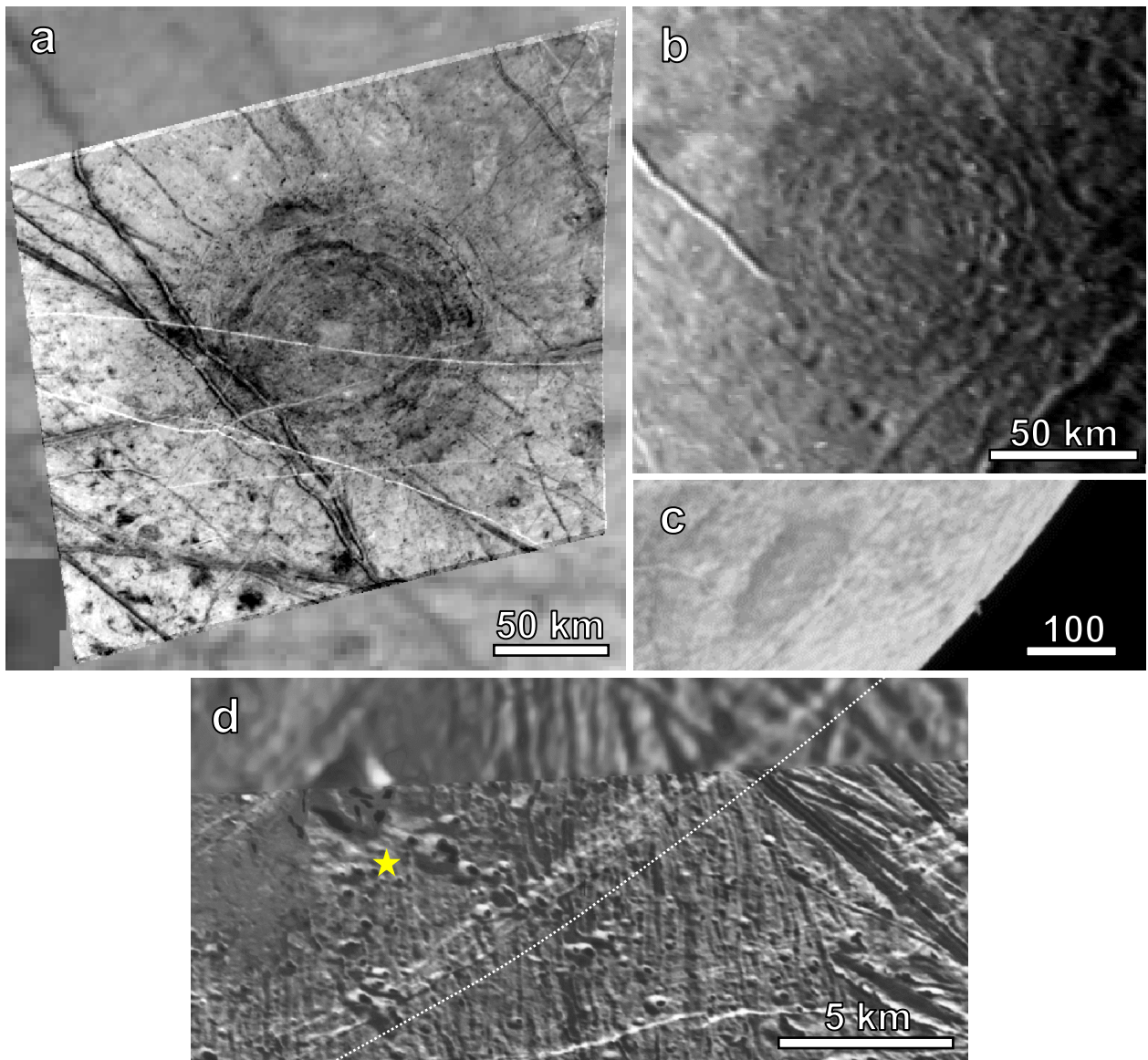

**Figure 4. Additional views of Tyre and Callanish.** (a) High sun Tyre image from *Galileo* Mosaic G7ESTYRMAC01 (~590 m px−1) with G7ESGLOBAL01 (3.2 km px$^{-1}$) in the background (Bland et al., 2021). Dark material is associated with some of the graben. (b) Highest resolution full view of Callanish from *Galileo* mosaic 25ESGLOBAL01 (955 m px$^{-1}$). (c) High-sun Junocam view of Callanish near the limb of Europa from a raw, un-projected image (red filter) during the Europa flyby on 29 September 2022 (~1 km px$^{-1}$; https://www.missionjuno.swri.edu/junocam/processing?id=JNCE_2022272_45C00001_V01).
(d) Portion of high-resolution image mosaic (~30 m px$^{-1}$) near Tyre that overlaps with a small portion of a shallow outermost graben. Note the sunlight is from the lower left and this image is higher sun than the full mosaic covering Tyre. Dashed outline shows the general arc of the graben but lies outside and to the right of it so as to not cover the feature. Yellow star indicates matching position in Figure 2a. Aligned secondary craters overly the graben structure at left.

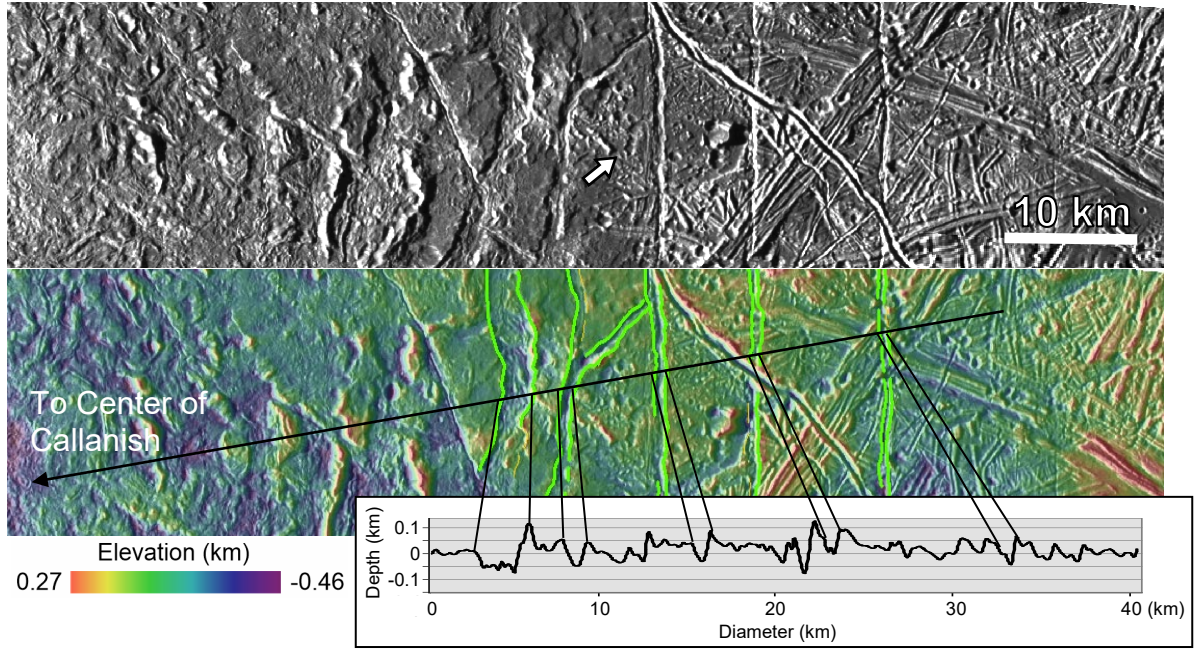

**Figure 5. High-resolution topography of the eastern portion of Callanish.** Prominent graben as shown in both the base image (45 m px$^{-1}$) and topography. Note 10-km-wide area of chaos overprinted by the narrow ring-graben at center (white arrow).

## 2.2. Measurement Techniques

### 2.2.1 Graben Morphology and Identification

Moving radially outward from the center of both Tyre and Callanish, there is a zone of disordered and rugged terrain that transitions into more ordered concentric structures. The very center-most portions of the structures have the lowest roughness amplitude, but are still composed of what appear to be blocky, hummocky terrain, especially with increasing distance from the center of symmetry. The inner circumferential structures are more ridge-like but some could be tilted blocks or half-graben. The distinctive circumferential ring-graben are pairs of what are or appear to be parallel normal fault scarps bounding a down-dropped block/central valley (bold green lines in **Figures 6-7**). They begin at ~45–50 km out from the center of Tyre and ~35–40 km from the center of Callanish, which is ~1.1-1.3 times the effective diameter estimates above. A few circumferential structures are less clearly graben, either because they are shallow, or because one bounding scarp is more obvious than the other (thinner orange lines in **Figures 6-7**). The graben become shallower and less wide at the structure outer edges (this can be seen visually and in the plots presented below). Some graben are more continuous at a given distance from the structure center, while others occur over only a short circumferential arc. There are up to five distinct graben for a given radial direction. For the below analysis we used all tectonic structures that were circumferential and displayed two bounding fault walls with an apparently down-dropped surface between them.

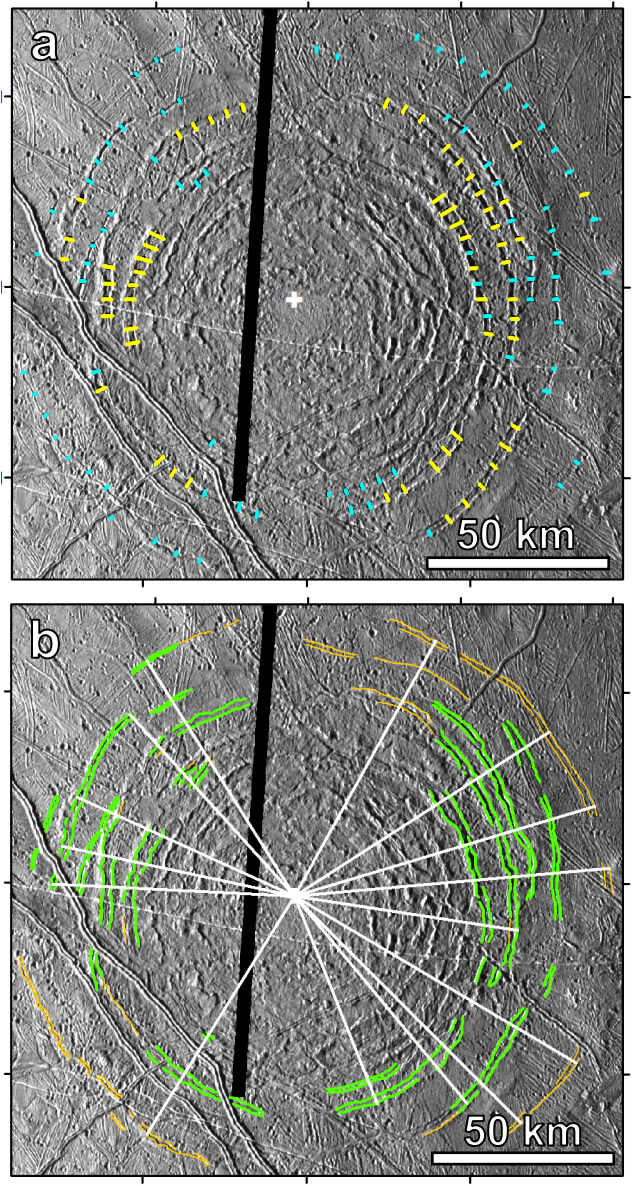

**Figure 6. Tyre graben mapping and measurement.** (a) Graben widths were measured at 5° intervals. Lighter, yellow lines indicate graben width that are wider than the average value (1.8 km) and darker blue lines indicate those smaller than the average. (b) Graben as mapped for this study (same as in Figure 2c). Solid white lines indicate the radial profiles used to measure graben depths and radial strain. All panels have the same geographic extent as in Figure 2..

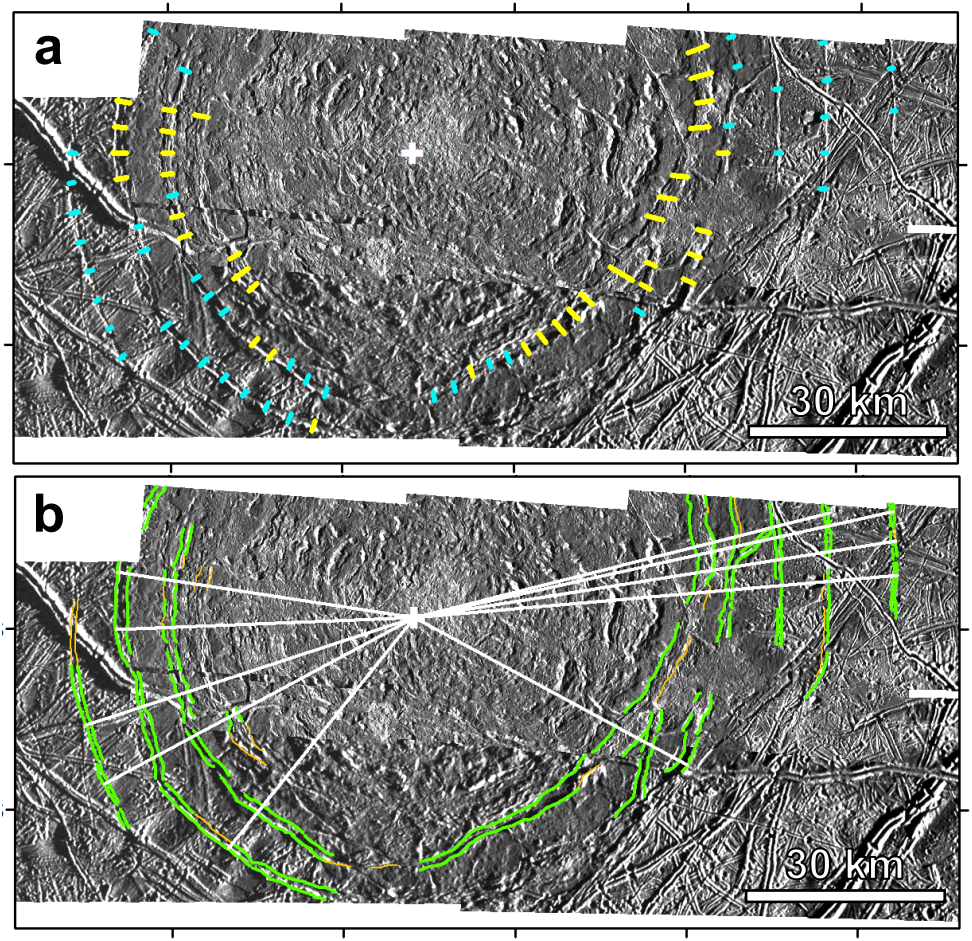

**Figure 7. Callanish graben mapping and measurement.** (a) Graben widths were measured at 5° intervals. Lighter, yellow lines indicate graben width that are wider than the average value (1.4 km) and darker blue lines indicate those smaller than the average. (b) Graben as mapped for this study. Thicker green outlines indicate more prominent graben walls, while thinner gold outlines indicate more subtle features or features with only one apparent scarp. White plus is the fitted center of the basin, and solid white lines indicate the radial profiles used to measure graben depths and radial strain. All panels have the same extent as shown in Figure 3.

In **Figure 8a** we show a closer view of the outermost trough/graben of Tyre to the NE. This marks the observed limit of impact-related tectonic deformation in this direction, and is notable for the topographic subtlety of the feature (it is only ~30 m deep) and for being fully illuminated, as opposed to often shadowed troughs elsewhere at Tyre. The preexisting structural "grain" of the background lineated plains are clearly present, i.e., downdropped. This is consistent with a graben structural interpretation, even though the resolution here (~0.2 km px$^{-1}$) is too poor to clearly identify footwall scarps. Such downdropping is inconsistent with, e.g., interpretations of these troughs as fissures (mode I tension cracks) subsequently filled by icy regolith or other debris. This is reinforced by the highest-resolution (30 m px$^{-1}$) view of any of the europan ring-graben, at southeast Tyre (**Figure 4d**). This high-sun image (incidence angle 30º) exhibits narrow, bright lineaments, which we interpret as illuminated slopes corresponding to the bounding graben faults or fault breaks, as well as additional, parallel tectonic elements (fractures or subsidiary normal faults) on the graben floor.

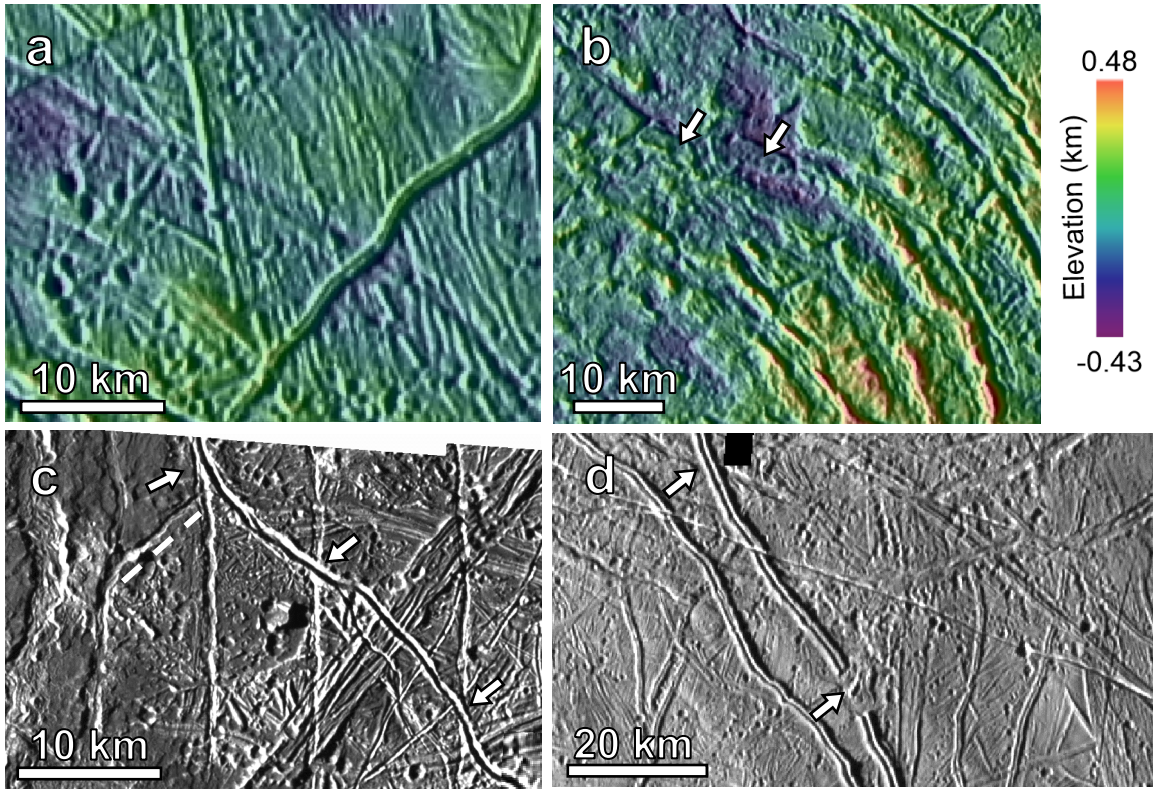

**Figure 8. Graben transition zones and ridge/chaos superposition relationships.** (a) The outermost trough of Tyre to the NE is ~1.5 km wide and ~30 m deep (see Figure 2b, upper right). Preexisting lineated plains are clearly downdropped within the (unshadowed) troughs, consistent with a graben origin. (b) Transition between Tyre's NE graben zone (upper right) and annular massif zone (lower left). Note apparent in-warping of ring geometry at center. Oblong (~8 km) crustal blocks (horst fragments?) have apparently shifted (rafted?) inward by up to several kilometers (arrows). (c) Eastern Callanish graben (Figure 3b,c, upper right) crossed by later double ridge (arrows) and example linking fault between two graben (next to dashed line). (d) Tyre graben crossed by later ridge, that itself is partially destroyed by later chaos (region just to the south of the outer Tyre graben, extending beyond the frame in Figure 2). The lighting is from the left in all images.

Secondary craters are visible amongst the graben, and begin to occur at a similar radial distance from the crater center as the start of the more distinct graben (outside of the central rougher disordered terrain and more ridge-like concentric structures, e.g., as shown in **Figures 2, 3, 4d, 5, 8a,c,d**). The secondary craters are more obvious between the inner graben for Tyre (Moore et al., 2001; Singer et al., 2013a), whereas the Callanish secondary craters only become more apparent between the outer graben. It has been previously noted that Callanish secondary craters are not as prominent as around some other europan craters (Moore et al., 1998), but again we are only able to see part of the structure, and the lighting is fairly oblique. We return to the relationship between graben and secondary craters below (in Section 4.3) as a possible constraint on the timing of graben formation.

During mapping we also kept in mind that there may be pre-existing ridge or band structures that could potentially be conflated with graben. A visual examination shows that there are only a few visible pre-existing ridge or band structures that happened to strike in a similar manner to the circumferential graben. We did not include these structures in the measurements. In general, only a few ridges cross the ring-graben or other areas of the structures. A few later double ridges superpose the southwest portion of Tyre's ring-graben (**Figure 2a**), and one young double ridge transitions to more of a fracture or fissure as it runs through the graben and disrupted area of Tyre, emerging to the east as a simple fissure (Kadel et al., 2000). This ridge morphology change could be influenced by the ice shell material properties within the impact structure, but as one of the two stratigraphically youngest ridges mapped by Kadel et al. (2000) is just as likely the result of an incomplete morphological transformation or evolution (fracture to double ridge). There is one double ridge that crosses some of the eastern graben of Callanish (**Figure 8c**). There are no very obvious cases of graben being superposed by later chaos at Callanish. At the very southern extent of Tyre there is one chaotic patch that does interrupt a prominent post-Tyre double ridge and potentially could have obscured any graben structures there (**Figure 8d**).

The structure center points were determined by least squares fits to the radius of curvature of the visible graben or concentric ridge arcs (Lichtenberg et al., 2006). The center points determined this way do not fall directly in the middle of the central circles of less-rough terrain. For Tyre, the measured graben distances for the east and west sides (see below) were on average similar. For Callanish, the eastern and western sets turned out somewhat offset from each other, which may indicate that center location was not as well estimated because only half of the structure was imaged, or the impact structure may in fact be asymmetrical. There is a smaller (~10-km wide, 100-m deep) depression in the chaotic central area of Callanish (Schenk & Turtle, 2009), but it is also offset from the visual center of the graben arcs (see their Fig. 11). For the results below, we use a radial distance (range) for the maximum strains, and this would not be strongly affected by shifting the Callanish center point, as the sets would still average out to produce similar results (one set would shift closer to the center and one would shift farther away, and the shift would likely only be ~5 km in many cases). Accordingly, we decided to use these center points, which were fit the same way for both impact structures based on the graben arcs as described above, rather than visually selecting the center of Callanish.

### 2.2.2 Graben Measurements

All measurements were made along transects radial to the fitted center of the structure. Graben widths were measured in equal 5° intervals wherever two (inferred) graben walls were crossed by a radial transect line (**Figures 6a and 7a**). Graben spacing was measured between the center points of adjacent graben on the same 5° radial lines. Graben depths were measured along radial topographic profiles that were selected specifically to cross multiple graben (**Figures 6b and 7b**), while also attempting to sample at intervals around the structures (and do occasionally fall on the same azimuths as some of the 5° radiants).

In some cases, the graben floor was resolved in the topography as a wider, flatter area (e.g., **Figure 5**), and in other cases the graben profile was more u- or v-shaped. The height for both the inner and outer (inferred) fault scarp of each graben was measured using the high point at the top of the scarp wall (see examples in **Figure 5**), and the nearby elevation at what appeared to be the base of the exposed scarp at the graben floor. Both the lighting geometry and the physical geometry of the graben themselves sometimes produced shadows that obscured a portion or all of the graben floor. The photoclinometry technique cannot estimate elevation values inside shadowed regions. Instead, the incidence angle is used to effectively trace the "tops" of the shadowed regions as a height estimate and to create a continuous topographic product. If the shadow falls onto the graben floor this would still represent a similar measurement to an un-shadowed graben. However, if the shadow covers the entire floor and falls partially onto the opposite wall, this would represent a minimum depth, because the full depth would not be captured. Shadows likely affected some of the depth measurements at both Tyre and Callanish, particularly for the outer narrow graben (**Figures 2-3**). If the depth measurements are a minimum, that propagates to a minimum displacement and radial strain.

## 3. Measurement Results

The below sections describe in detail the results for each type of measurement, and a summary is provided in **Table 1**.

### 3.1. Graben Widths

Tyre graben are wider on average than Callanish graben, but both show a wide range of values, and a wide range of values at any given distance from the structure center (**Figure 9**). The running average for graben widths decreases with increasing distance from the fitted center, as can also be seen visually in the images. The minimum graben width (for a given distance from the structure centers) is unlikely to be a function of image resolution in most cases, as we are generally measuring distinct graben and not narrow fractures. However, the very outermost graben can become fairly narrow (at least for Callanish). The narrowest feature measured is one of the outermost graben for Callanish, as seen in the high-resolution eastern mosaic (**Figure 5**). The narrower spread in width values for the outer Callanish graben may be due to the fact that there is considerably less radial coverage than for the inner Callanish graben, but in any case there is no intrinsic reason why the graben width distributions for Tyre and Callanish should be the same. The primary control is lithospheric/thermal structure at the time of impact, as will be delved into detail in Section 4. For both structures, we use the overall average graben width for many of the calculations below, but also address the running average as a function of radial distance as well as the range of graben widths, which as noted is decidedly broad (**Figure 9b**).

## 3.2. Graben Depths

As described above, the depths were measured for each graben wall. For **Figure 10,** we plot the average depth (average of both inner and outer wall measurements) for a given location along the white radial lines shown in **Figures 6b and 7b**. Tyre graben are slightly deeper on average than Callanish graben. The shallowest graben measured in this study were measured in the higher resolution mosaic/topography for Callanish (45 m px$^{-1}$) and are about 15 m in depth. The lower bound for depth measurements is nearing the vertical precision of the topography, but even Tyre's shallow outer graben (~40–50 m on average) are still resolved topographically. Overall, we do not find a preference for the inner or outer graben walls to be systematically deeper or shallower than the other. For eastern Callanish, however, there is a preference for the outer wall of the inner graben to be deeper (possess greater throws) (**Figure S1** in Supporting Information). Future higher-resolution images should be able to clearly determine if either of the bounding faults of a given graben is the primary.

## 3.3. Inter-Graben Spacing

Average inter-graben spacing (for all graben measured) is larger for Tyre than for Callanish (**Figure 11a**). There is a slight-to modest increase in the running average of the spacing with increasing distance from the center of both structures (greater for Tyre); however, there is generally a wide range of values at any given distance (**Figure 11b**). This wide range of values is largely due to azimuthal variations; in a given azimuthal sector the progressive increase in inter-graben spacing with radial distance is clear in map view (**Figure S2** in Supporting Information). This general increase in spacing is a well-known characteristic of impact ring spacing, though far from the classic $\sqrt{2}$ ring ratio oft-quoted for lunar basins (Melosh, 1989). There is less of a spread for the Callanish graben spacing values, and there the outer graben especially cluster more tightly than for Tyre, which again may be due to the lack of radial (if not azimuthal) image coverage for Callanish. The variation in graben spacing around the structures illustrates how the graben are often not continuous around the entire structure, and there are a few locations where linkages may have occurred, leading to short spacings as some of the faults merge or nearly-merge (some examples are the complex region in the SE portion of Tyre and also the linking fault shown in **Figure 8c**). The innermost recognizable graben structures occur at ~1.9 and ~2.2 times the estimated effective crater rim radius out from the center of Callanish and Tyre, respectively. The farthest recognizable graben are ~4.5 times the transient crater radius away from the center of each structure, but this value may be somewhat limited by image resolution (for Tyre) or coverage (for Callanish). As previously described, between the effective rim and the first appearance of circumferential graben for each structure lies a zone of extremely rugged arcuate ridges, annular massifs, blocks, and other materials resembling chaos units elsewhere on Europa (Moore et al., 1998; Moore et al., 2001).

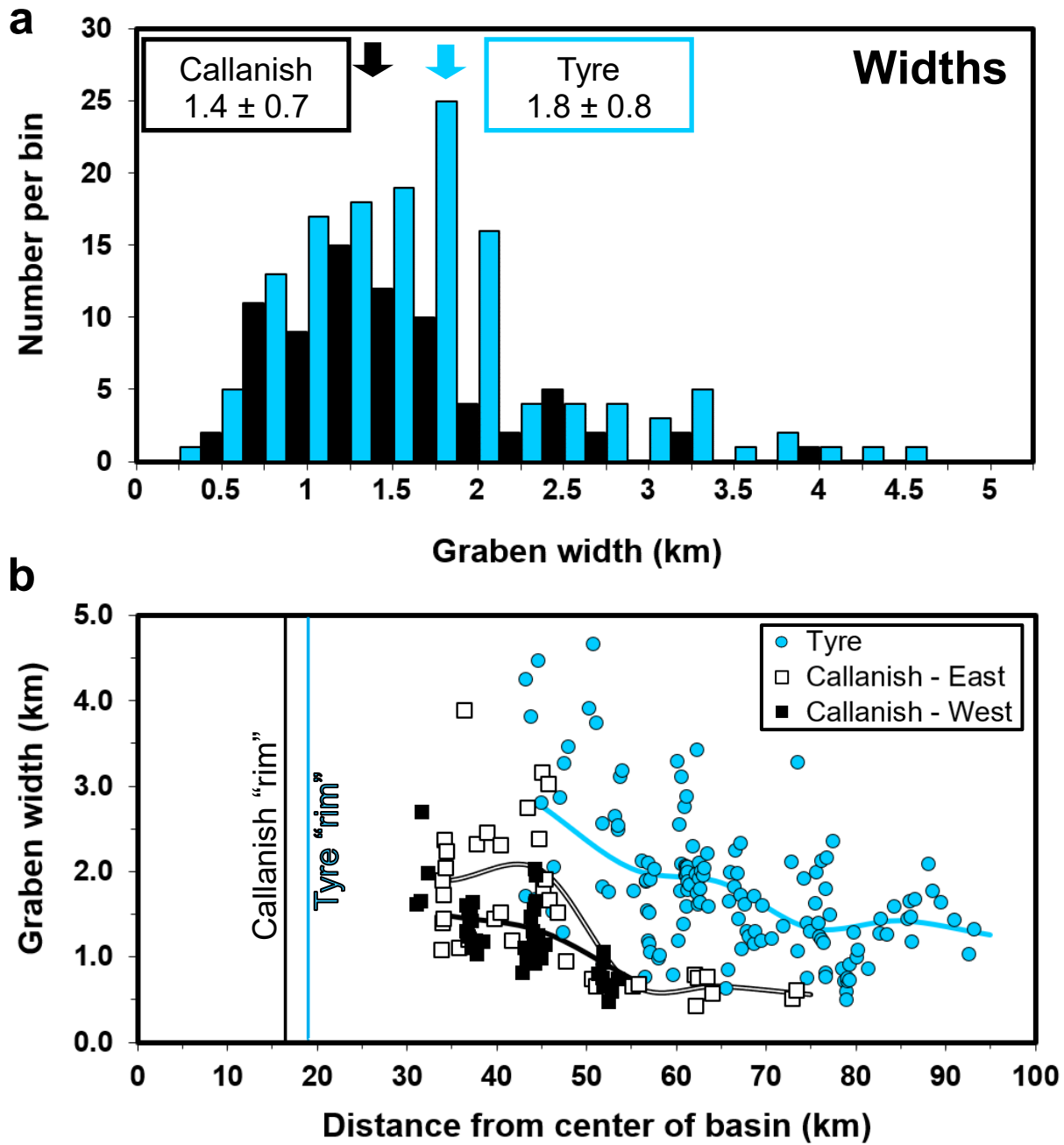

**Figure 9. Graben width measurements.** (a) Histogram of graben widths and average values for both basins. (b) Graben widths as a function of distance from the basin center with each basin equivalent rim radius indicated for reference (Table 1). Distance is given from the center of the structure to the center of each measured graben. Curves show the running average widths over a 10-km distance bin (box average). Graben widths generally decrease with increasing distance.

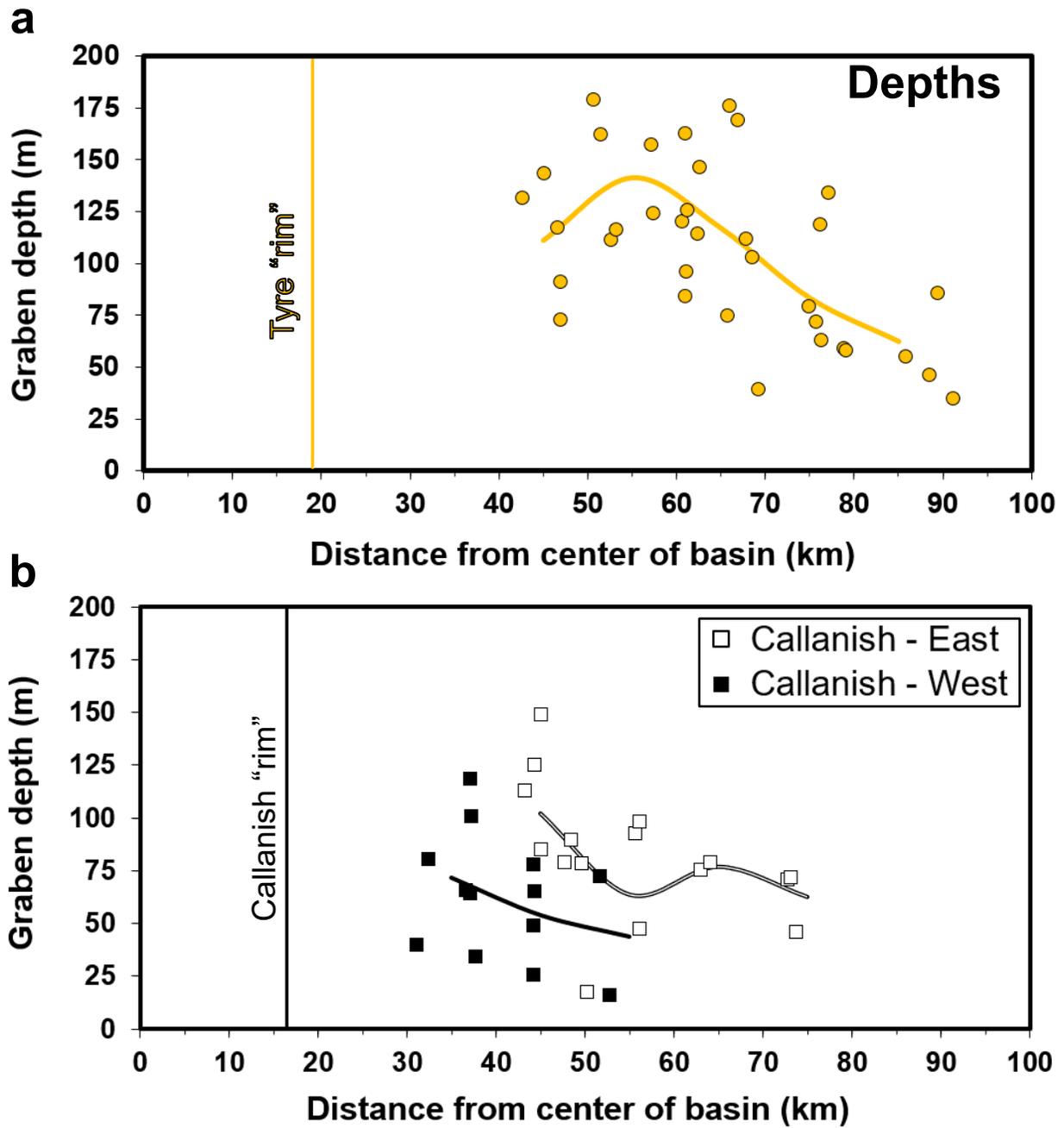

**Figure 10. Graben depth measurements.** (a) Tyre graben depths and (b) Callanish graben depths as a function of distance from our calculated basin center for each structure to the center of the graben. The basin equivalent rim radii are indicated for reference. Curves show the running average depths over a 10-km distance bin (box average). Graben depths generally decrease with increasing distance.

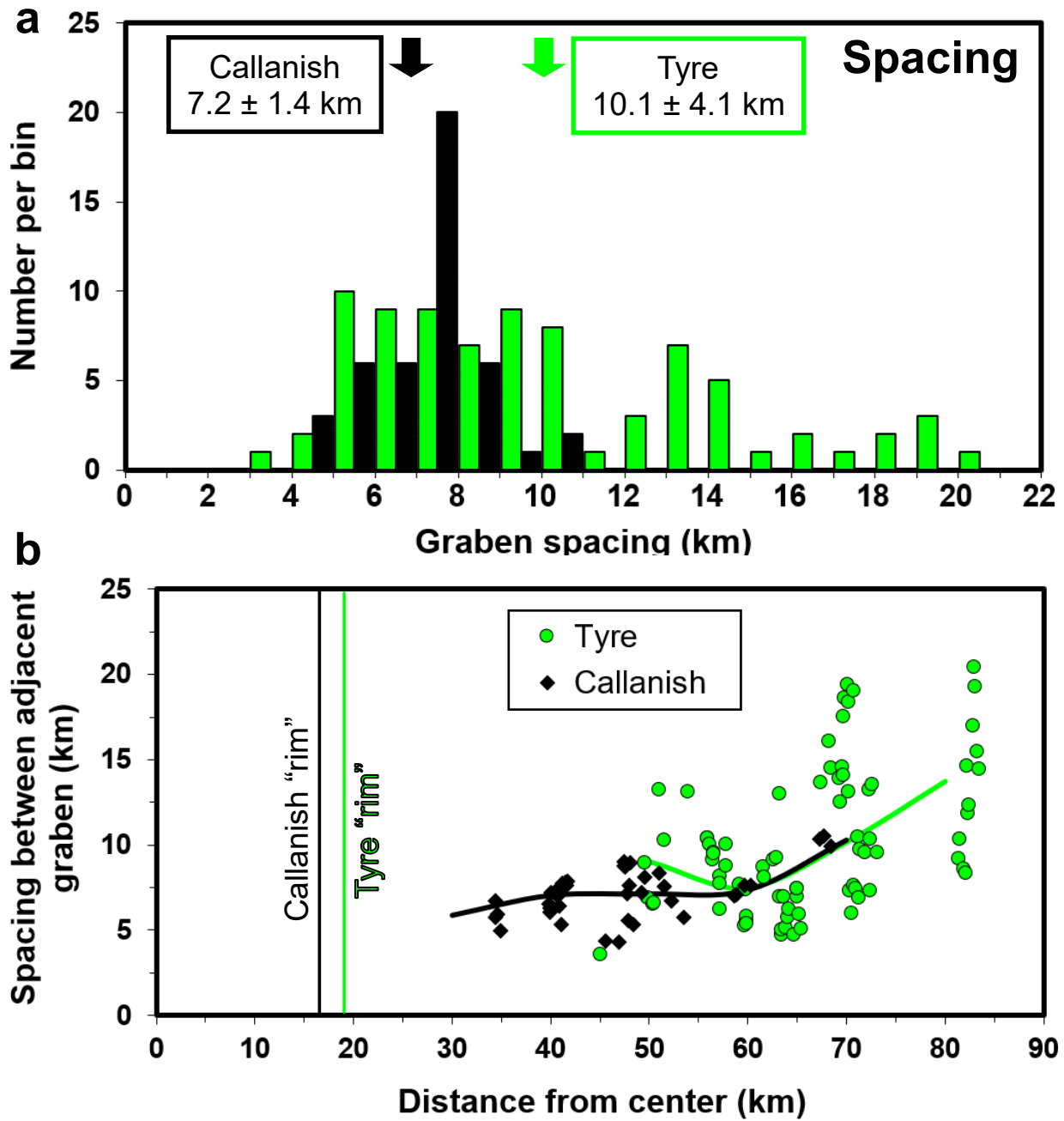

**Figure 11. Graben spacing measurements.** (a) Histogram of graben spacings and average values for both basins. (b) Graben spacings as a function of distance from the basin center with the basin equivalent rim radius indicated for reference. Graben spacings distances (x-axis on panel b) are measured from the center of Callanish or Tyre to the mid-point between the graben walls (also see Figure S2). Curves show the running average widths over a 10-km distance bin. Graben spacings generally increase with increasing distance.

**Table 1. Tyre and Callanish Morphometry**

|  | Tyre | Callanish |
|---|---|---|
| Estimated equivalent rim diameter[*] | 38 km | 33 km |
| Estimated transient diameter[†] | 23 km | 20 km |
| Estimated impactor diameter[⁕] | 1.74 km | 1.45 km |
| Farthest radial graben extent (from structure center) | ~95 km | ~75 km |
| Graben width – Range | 0.5–4.7 km | 0.4–3.9 km |
| Graben width – Average of all measured[‡] | 1.8 ± 0.8 km | 1.4 ± 0.7 km |
| Graben depths – Range | 30–210 m | 14–150 m[§] |
| Graben depths – Average of all measured[‡] | 107 ± 41 m | 75 ± 35 |
| Graben spacing – Range | 3.6–20.5 km | 4.3–10.5 km |
| Graben spacing – Average of all measured[‡] | 10.1 ± 4.1 km | 7.2 ± 1.4 km |

[*]Based on scaling from secondary craters, see description in Introduction.
[†]Based on scaling in McKinnon & Schenk (1995).
[⁕]Based on scaling in Singer et al. (2013a), Equation 10.
[‡]Variance reported as one standard deviation.
[§]Includes both the lower resolution western set and the higher resolution eastern set.

## 4. Analysis and Interpretations

### 4.1. Radial Displacement and Strain

  **Figure 12** describes the flow of information that leads to the results described here, from the mapping and measurements at Tyre and Callanish to inferences regarding impact ring formation and properties of Europa's ice shell, at least at the time of impact. One of the first measures to address concerns the integrated nature of the deformation (strain) implied by the formation of the outer graben zone in each case. Obviously, a given circumferential graben implies a radial extensional stress environment and resulting radial strain due to normal fault displacement. The radial displacement ($\Delta r$) per graben wall is calculated through simple trigonometry (equation below) using the depth (throw) on each side of the graben and an assumption for the fault dip angle (**Figure 13a**). A simple graben geometry is assumed, with steeply dipping planar normal faults (not curved or listric), and we ignore any possible shallowing by mass wasted ice debris. The latter is difficult to assess given the resolution of the Tyre images, but we note that at least the shallow troughs/graben preserve their lineated plains topography in their downdropped (hanging-wall) sections (**Figure 8a**).

  Accordingly, the dip angle ($\theta$) of the normal faults should be equal to $\pi/2 + \varphi/2$, where $\varphi$ is the friction angle, following classic Andersonian fault mechanics (e.g., Jaeger & Cook, 1979). We used $\varphi = 34.6°$ as found in the laboratory experiments of Beeman et al. (1988), from their low-pressure coefficient of friction $\mu = 0.69$ (when assuming zero strength at zero normal stress), which yields a dip angle of $\theta = 62.3°$. For each wall, $\Delta r = d/\tan\theta = d/1.905 \approx d/2$. Thus, when adding both walls together, the total radial displacement per graben is approximately equal to the average depth of each graben. Inward radial displacements per graben are in the range of 50–160 m and 30–105 m for Tyre and Callanish, respectively. Very similar measures are found if we adopt the Beeman et al. (1988) recommended $\mu = 0.55$ ($\varphi = 28.8°$ and $\theta = 59.4°$), which they note may have been affected by the strength of the indium jacket used in their ice friction experiments. We note, however, that more complex fault geometries are possible, based on terrestrial field analogues, which will be discussed in Section 4.5.1.

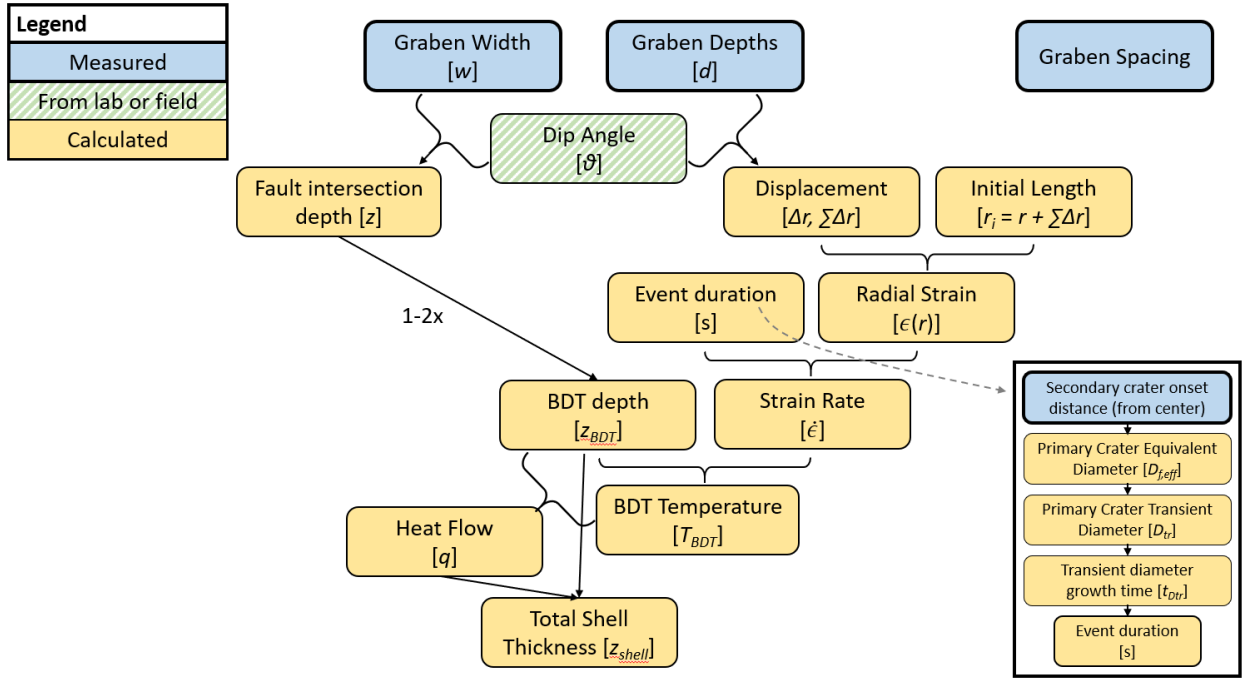

**Figure 12. Flow of information for estimating BDT (brittle-ductile transition) depth, heat flow, and total shell thickness.** The top row of values in the blue boxes with bold outlines are quantities measure in this study. The dip angle is an assumed value derived from laboratory measurements of friction angle. The remaining values (yellow boxes) are calculated in this study.

The total radial strain, $\varepsilon(r)$, experienced by the impact structure terrain is necessarily lowest among the outer graben (asymptoting to zero at large radial distance $r$), and accumulates with each successive graben to higher values toward the center of the basin where the original surface has been displaced inward to a greater extent. Overlaid on this is a pattern of shallower graben with decreased widths and increased spacing in the outer portions of both structures (these smaller displacements likely related to the asthenospheric strain distribution and fault formation sequence, as will be discussed). To understand the typical maximum values of accumulated radial strain we incrementally sum the strains for each graben along each of our radial profiles from the outside in (**Figure 13b**). We divide by the estimated initial length for each point along each radial profile ($r_i = r + \sum \Delta r$) to calculate the strain. The initial length is the estimated original radial distance from the center of the structure to each surface point along the profile before the displacement occurred.

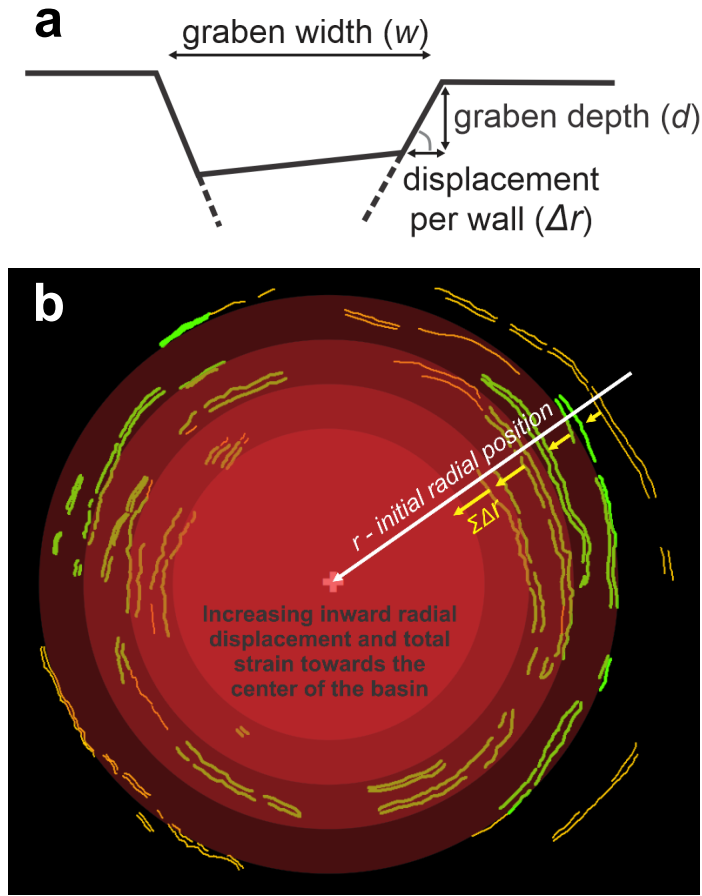

**Figure 13. Measurement geometry.** (a) Cartoon of graben geometry. Small gray arc indicates the location of the angle $\theta$. (b) Illustration of how the cumulative radial displacement ($\sum \Delta r$) and strain ($\varepsilon$) increases from the outer to the inner areas of the basin.

We plot the radial strain, $\varepsilon(r) = \sum \Delta r / r_i$, as a function of the current radial distance from the center of each structure ($r$) in **Figure 14**. The typical maximum values of strain for Tyre are about 1–1.5% and for Callanish are ~0.6–1.2%. The total value of strain for each radial profile will vary as the profiles cross varying numbers of measurable graben. Values of ~1% extensional radial strain for the inner graben imply similar contractional circumferential strains, and in the absence of additional, secondary faulting, circumferential lithospheric stresses (i.e., those parallel to the graben strike) of order $\varepsilon \times E \approx 100$ MPa, for a solid ice Young's modulus $E$ of 9.3 GPa (Petrenko & Whitworth, 1999). Such stresses are arguably more compatible with the prompt collapse of large impact transient cavity than with long-term isostatic adjustment, but the more important point is that such stresses are simply not sustainable in Europa's ice lithosphere: they are far larger than the strength of ice, even in compression (Arakawa & Maeno, 1997; chapter 11 in Schulson & Duval, 2009), and further deformation (faulting) is inevitable. (For an ice lithosphere riddled with myriad faults both $E$ and the compressional stresses implied would be lower, but the yield strength of such a lithosphere would be less as well). Such stresses, or their ramp up, are likely responsible for the rugged topography (annular massifs, displaced and possibly rotated crustal blocks) inward of the graben zone of each structure. For example, **Figure 8b** shows a close-up of the Tyre DEM in which apparent fragments of inter-graben ice shell have broken along fractures oblique to the radial direction and been displaced inwards, by possibly several kilometers in one case. Strains locally have clearly well exceeded 1%.

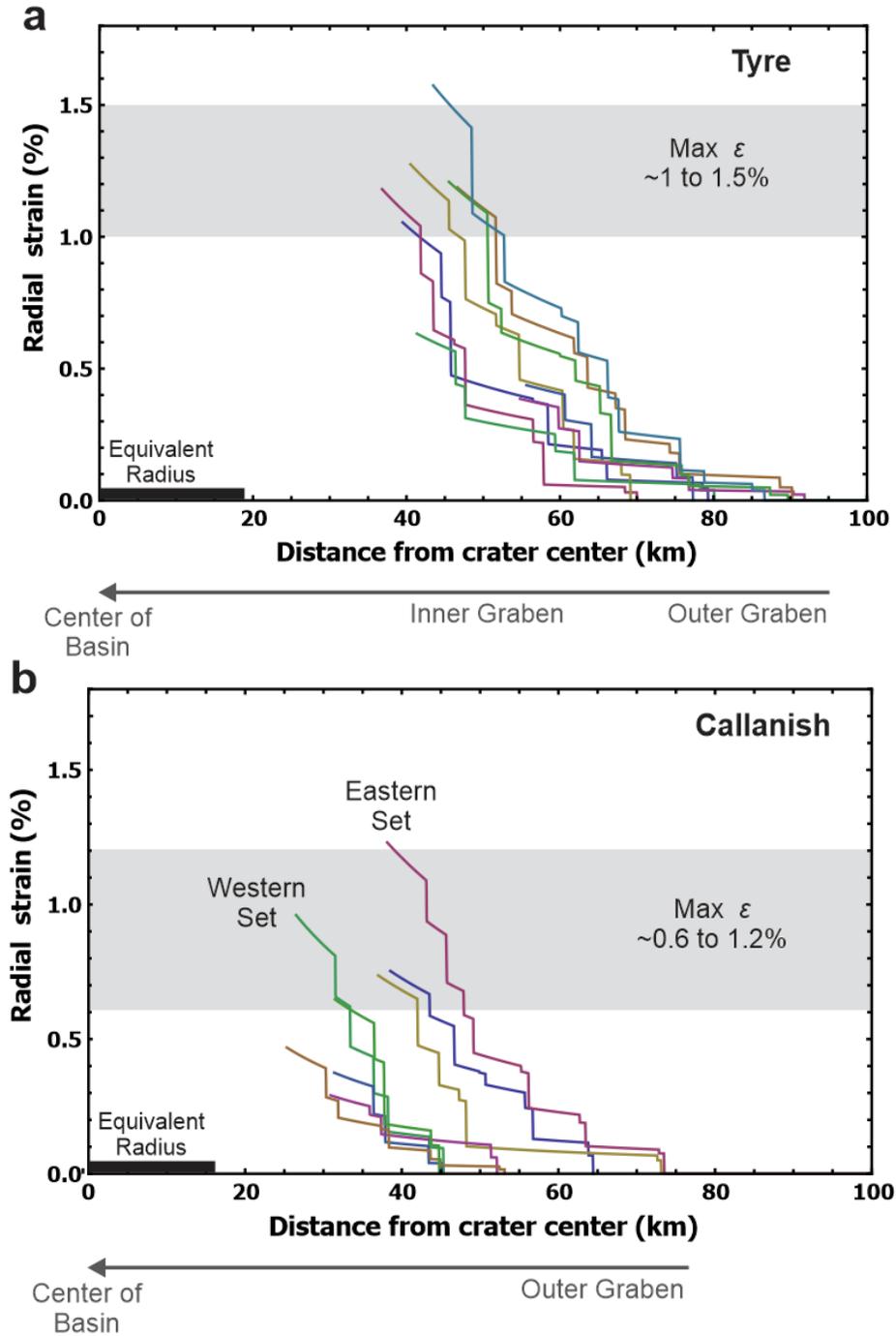

**Figure 14. Cumulative radial strain.** Starting with the outer graben and moving inward, we sum the measured strains for each basin along the profiles shown in Figure 6-7b.

## 4.2. General Morphologic Constraints on Ring Tectonics

The striking appearance of Tyre and Callanish and their similarities to the much larger Valhalla-class multiring basins on Callisto have long prompted discussions of thin ice lithospheres and asthenospheric inflow as proposed by McKinnon & Melosh (1980). In the initial *Galileo* analysis of Moore et al. (1998), it was concluded that an ice shell ~10-to-15 km thick overlying liquid water could account for both the ring graben surrounding Callanish and Tyre *and* preserve the defined rim and central peak of the somewhat smaller crater Pwyll (24-km final diameter). Subsequent to the receipt of the high-resolution Tyre mosaic, this estimate was refined to a likely maximum thickness of the brittle(-elastic) ice lithosphere of $10 \pm 1.5$ km, but with a preferred brittle lithosphere thickness of $3.5 \pm 0.5$ km (Kadel et al., 2000). These latter authors were agnostic as to whether the subjacent "asthenosphere" (weak substrate) was warm ductile ice, intra-crustal brine, or the ocean itself at the time of impact, though Moore et al. (2001) suggested that ~10-11 km was a plausible depth to fluid (meaning water or brine) beneath both impact sites.

Part of the thinking behind these limits was the (at the time) understanding of ring formation as a function of lithosphere thickness as proposed in McKinnon & Melosh (1980). For example, in Fig. 3a in that work, the violent disruption of a very thin ice shell was shown, the general idea being that for a very thin shell over an ocean a series of large amplitude gravity waves (tsunamis) would propagate outward and the ice shell would have little hope of not being fragmented. Modern numerical shock physics calculations of large impacts into Europa's ice shell are, during excavation, not limited by strength, and the transient depths reached are quite large, ~0.5 × the transient diameter (Turtle & Ivanov, 2002; Bray et al., 2014; Cox & Bauer, 2015; Silber & Johnson, 2017). If this depth approaches or breaches the ice shell (the total thickness), multiple oscillations of a highly fluid, water-rich central peak are unavoidable, and it seems likely that the ice shell just beyond the transient rim would be not only strongly shock damaged but suffer extreme disruption and rearrangement during the dynamic collapse phase. It seems highly plausible to us that the rough inner unit of Tyre corresponds closely to this (qualitatively predicted) disrupted region, especially as it would also have been dynamically overlain with ice-slurry ejecta, which would further subdue its final appearance. In this spirit our understanding of the effective rim diameters quoted here and in Schenk & Turtle (2009) can be refined: there never was a final rim in a structural sense, and $D_{f,eff}$ values only serve as fiducial markers and for comparative purposes. Transient depths and diameters are, however, quite real (or were, if only for an instant).

Following from the above, the outermost annular region of tectonic failure, the graben zone, does not respond to propagating tsunami waves in Europa's ocean, but more to the time-integrated inflow at depth refilling the transient cavity, essentially to its original, pre-impact elevation (McKinnon & Melosh, 1980). The concept we are advocating is that impacts into a suitably thin icy shell can create, beyond the transient cavity, *both* an inner zone of extreme disruption and an outer zone of ring graben (and apparently, an intermediate zone of annular massifs that shares characteristics of both).

**4.3. Constraints on Graben Formation Timescales From Superposition Relationships**

Previous geologic mapping for Callanish suggested that the superposition relationship between the interior continuous ejecta deposits and the ring-graben indicated they likely formed closely in time (Moore et al., 1998; Moore et al., 2001). Given the general rough nature of the europan surface and of the structures, the extent of continuous ejecta is somewhat difficult to discern, although it is generally associated with a lower albedo in high sun images (Moore et al., 1998; Schenk & Ridolfi, 2002; Schenk & Turtle, 2009). In the case of Callanish, some ejecta was considered to be cut by the graben, and some appear to be superposing graben (Moore et al., 2001, their Figure 10), indicating the close timing.

We also consider any constraints on timing of the formation of the graben after the impact that might be gathered from the relationship of the graben with secondary craters. As noted earlier, secondary craters are more apparent among the Tyre graben, whereas the secondary craters around Callanish seem to pick up in density only outside of its graben and are not as abundant. The largest secondary craters in the Tyre area are larger than some of the graben widths, but the image resolution for the main mosaic covering Tyre is still somewhat marginal for determining superposition relationships. There is always the chance that a small crater in this region is instead a primary crater, but as was previously shown, even small primary craters are not abundant on Europa (Bierhaus et al., 2005), and the examples described below are parts of chains of appropriately sized craters and thus they are likely secondaries radial to Tyre.

Several observations that may bear on timing are: (1) In all of the *inner graben* floors that can be seen around Tyre (the ones that are not shadowed), there are no obvious secondary craters on them, while there are secondary craters on the terrain in-between graben **Figure 15a** and some that appear to come up to and stop at the edge of some inner graben**;** (2) Some secondary crater chains appear to overprint the *outer graben* **(Figures 4d and 15b)**. **Figure 4d**, from the high-resolution E15 sequence, is especially compelling in this regard; and (3) There are no clear cases of graben cutting secondary craters. Even in the high-resolution mosaic it is difficult to make the case that any of the secondary craters have been subsequently modified by graben.

A scenario that could explain these observations is that the secondaries formed after the graben, but the inner graben, which have greater displacements in general, were tectonically disruptive enough due to continued fault motion to effectively erase any previously formed superposed secondary craters. Using a 45° ejection and impact angle, and the distance of some of the innermost secondaries from a plausible launch location (~1/2 the transient crater diameter) of ~40 km, the launch velocity would have been ~250 m s$^{-1}$ and a travel time of ~4 min for a ballistic trajectory on a sphere (Singer et al., 2013a). Using the same parameters as above and a distance of ~60 km (to the outer graben), the launch velocity would have been ~280 m s$^{-1}$ and a travel time of ~5 min. This timescale is order-of-magnitude similar to the timescale for transient crater collapse (~2 min, see Section 4.5), so it may be reasonable for the inner graben to continue to expand and deepen after secondary-forming fragments land. Future higher resolution images from *Europa Clipper* should enable a definitive determination of the timing between secondary crater formation and graben formation as a function of radial distance.

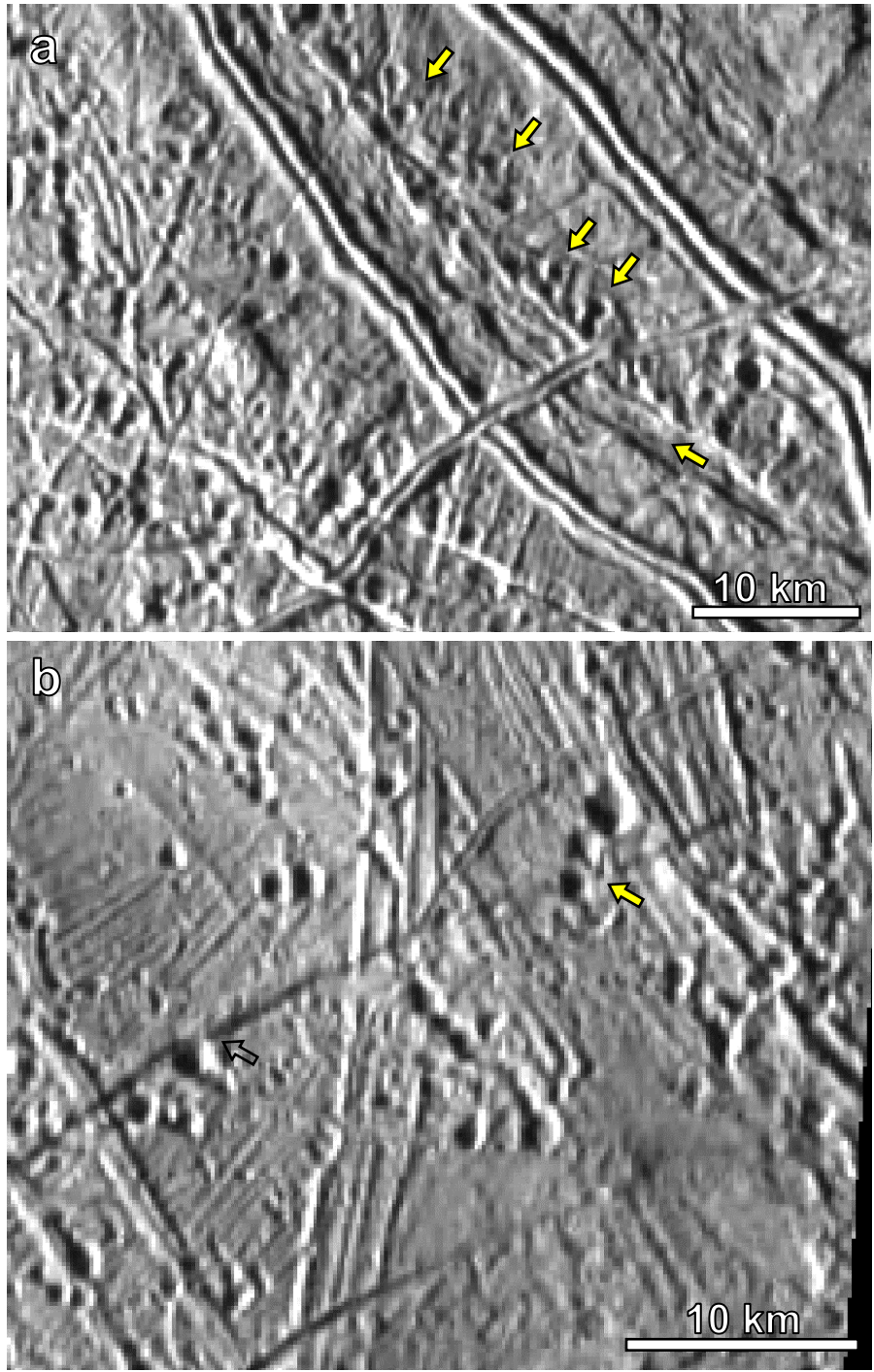

**Figure 15. Association of secondary craters with Tyre graben.** (a) An example area with secondary craters in the surrounding terrain but not apparent inside of a prominent, wide graben (lower arrow runs along strike inside of the graben and upper arrows point to secondary crater chains outside of graben). (b) Shallow outer graben in the northern part of Tyre, where a chains of secondaries craters (arrows point to intersection of graben and probable secondary craters) may overprint a graben. Alternatively, this outer graben was not pronounced enough to cause clear visible disruption of the secondary craters at this image pixel scale (170 m px$^{-1}$). See additional examples of secondaries overprinting a shallow outer graben in Figure 4d.

**4.4. Shell Thickness Constraints From Prior Modeling**

Initial constraints on shell thickness from Moore et al. (1998; 2001) and Kadel et al. (2000) were cited above. Turtle & Ivanov (2002) then determined from their impact simulations that Europa's ice shell would need to have been ≳10-15 km thick in order for Pwyll to retain a central peak. (This limit was also consistent with an earlier upper limit of >3-4 km calculated from the depth distribution of shock-produced melt [Turtle & Pierazzo, 2001]). Subsequent numerical crater formation studies attempted to match the depth/diameter data assembled by Schenk (2002) for Europa as a whole, especially the maximum observed crater depths (roll-over) at 8 km diameter. Bray et al. (2014) found that a linearly conductive ice shell of 7 km over an ocean best matched the 8-km rollover (whereas a 10 km thick shell did not), but the simulation depths plotted on the "shallower extreme" of the Schenk (2002) data cloud (see their Fig. 4). Cox & Bauer (2015) in contrast found a good fit overall for a 10-km thick shell over an ocean (meaning that the simulation depths fall entirely within the envelope containing the europan measurements; see their Fig. 5), using a very similar iSALE setup as Bray et al. (2014). Silber & Johnson (2017) then more-or-less matched the Schenk (2002) depth/diameter data with an ~8-km thick conductive ice shell (simulation depths now at the deeper end of the data cloud, see their Fig. 8a), again using iSALE. The point of these comparisons is not that one set of models is superior to another, for they all use the same numerical code and each subsequent study built upon the earlier. Rather, with complex numerical models the devil is always in the implementation details and parameter choices (see Silber & Johnson, 2017).

None of the above published numerical studies focused on explaining the detailed morphometry of Tyre or Callanish, but given the transient crater diameter estimates in **Table 1** (23 and 20 km, respectively), the implication is that both impacts breached Europa's ice shell, if conductive, whether during excavation or subsequently during crater collapse (Cox & Bauer, 2015). Moreover, the relatively lower albedo and reddish color of the two impacted regions (especially Callanish; **Figure 3**) is consistent with emplacement of ocean water in the ejecta, whose salts — halides or sulfates, hydrated or not — are able to radiation darken (e.g., Carlson et al., 2009; Hand & Carlson, 2015; Hibbitts et al., 2019; Trumbo et al., 2019). These structural and compositional inferences are closely consistent with initial *Galileo* interpretations (Moore et al., 1998; Moore et al., 2001).

The question then becomes what do the ring graben, particularly their depths and spacing, tell us about the ice shell during the impact, and are any such inferences consistent with the numerical impact modeling done to date? In that regard, Silber & Johnson (2017) also modeled impacts into a conductive ice shell overlying warm, ductile ice. They found that they could also explain the overall Schenk (2002) crater depth-diameter trend with conductive lid thicknesses in the 5-7 km range overlying a warm, convective, but solid asthenosphere (at 255-265 K); see their Figs. 8b,c. These latter models were less successful at matching the central characteristics of craters such as Pwyll, but we will also consider these rheological models in the context of the annular graben. We focus next on the depths of the ring graben, and what we can infer about the thermal structure of Europa's ice shell at the time of impact.

**4.5 Subsurface Fault and Thermal Structure**

Interpretation of planetary grabens in terms of subsurface interfaces, lithological or otherwise, goes back many years (e.g., McGill & Stromquist, 1979).  For symmetric graben, it is a natural assumption that two conjugate faults can nucleate at a mechanical interface, and simultaneously propagate upward to the surface.  For impact-related graben on Callisto and furrows on Ganymede (taken to be impact-derived), it was hypothesized that the brittle-ductile transition for water ice could serve as the appropriate interface (Golombek & Banerdt, 1986).  Alternatively, a transition to a convecting ice interior could be that interface (McKinnon, 1982; McKinnon & Parmentier, 1986).

While not incorrect in toto, this concept of impact graben formation needs to be reexamined (following Schultz et al., 2007).  Fault motion in a brittle layer resting on a ductile substrate is still fundamentally important.  In certain terrestrial tectonic settings this is certainly the case.  The arcuate Canyonlands graben in Utah are perhaps the classic example (McGill & Stromquist, 1979).  There a clastic sedimentary layer sequence (mainly sandstones and limestones) is actively extending downslope atop a bed of ductile evaporites (mainly halite, gypsum, and anhydrite) toward the Colorado River.  Detailed field study has revealed that in some places gravity sliding is at work, with the more ductile evaporite being sheared from the top, whereas elsewhere the evaporite is flowing underneath due to pressure gradients, Poiseuille-style, and thus dragging the overlying rocks (Kettermann et al., 2015).  The important points from this and other field work are, however: (1) the graben can be asymmetric, with a primary, structurally controlling fault and a  secondary, structurally related antithetic fault; (2) fault initiation begins in the weakest layer, the vertically jointed sandstones at the top (near-zero overburden) and propagates downward; (3) fault dips are thus near vertical at the top of the section, and only shallow to dips appropriate to shear failure below the sandstones; and (4) the primary and antithetic faults of a given graben do not, generally, meet at the top of the evaporite layer but at an intermediate, though lower, point in the sedimentary sequence (McGill et al., 2000; Schultz-Ela & Walsh, 2002).

Another terrestrial analogue site of direct relevance to Europa is the Silverpit structure in the North Sea, a 20-km-wide circular arrangement of buried graben and compressional ridges, hypothesized to be an impact crater of the multiring Valhalla type (Stewart & Allen, 2002; Stewart & Allen, 2005). (A possibly similar submarine structure has been reported off the coast of Brazil as well [Correia et al., 2005], and see also the Nadir crater off the West African coast [Nicholson et al., 2022]). No direct confirmatory evidence of an impact origin for Silverpit has been found to our knowledge (Koeberl & Reimold, 2004), and the impact interpretation has been rather controversial in the reflection seismology community (e.g., Underhill, 2004; Stewart & Allen, 2005; Thomson et al., 2005), but this is not critical to the structural interpretation. 3D reflection seismology clearly shows a set of arcuate graben, on the western side of Silverpit, at the top of a buried Cretaceous chalk layer.  The opposing faults of the individual graben sole into this layer at intermediate depths, and Stewart & Allen (2002; 2005) hypothesize that intercalated clay layers and calcareous sediments dewatered to create ductile detachments within the chalk.

Stewart & Allen (2005) actually favor formation of the graben over a relatively long timescale (thousands of years after the initial "impact") due to down-dip gravitational stresses. This would accord with the minor offsets and relay ramps between the graben, as seen in plan, similar to those in low-strain-rate graben systems such Canyonlands (Schultz-Ela & Walsh, 2002).

On Europa the upper, colder ice shell acts in a brittle manner allowing formation of faults over the short timescales associated with impact structure collapse and modification, as it must. At some depth, however, as temperatures increase, the ice shell behavior transitions to a more ductile state, which we refer to as the brittle-ductile transition (BDT). Traditionally, the intersection depth of facing graben faults has been related to the depth of the BDT through the keystone or "V" model, where the BDT occurs at or slightly above the fault intersection depth (Golombek & Banerdt, 1986; Allemand & Thomas, 1991; Ruiz, 2005) **(Figure 16a)**. Is this likely to be correct for Tyre or Callanish? Top-down initiation of sliding on a primary normal fault followed by formation of a secondary conjugate fault implies intersection *within* the brittle ice shell, perhaps at the halfway point as proposed by Scholz & Contreras (1998) for terrestrial continental rifts. This is the hourglass or "X" model of graben geometry (Schultz et al., 2007) **(Figure 16b)**.

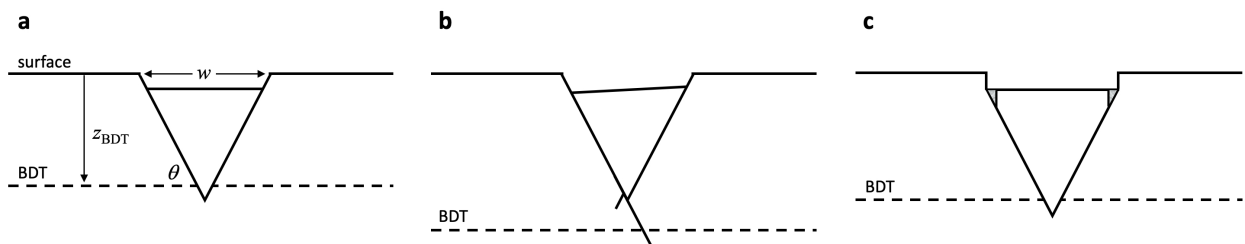

**Figure 16. Schematic graben geometries.** (a) Keystone or "V" configuration; (b) Hourglass or "X" configuration; (c) Similar to panel a, but with steep faults near the surface and minor infill by icy debris (shaded). Note that the BDT is inferred to occur at different depths for the same observed surface geometry (same $w$ and average graben depth), and that at least one fault intersects the BDT in all cases.

Numerical models of extending icy lithospheres employing strain weakening show "fault" (numerically, narrow zones of shear and plastic strain that serve as proxies for faults) intersection depths well within the upper half of the ice lithosphere (Bland et al., 2010), whereas an alternate, and arguably more physically realistic localization model shows intersections much closer to (if not at the *base* of) the lithosphere (Bland & McKinnon, 2015), at least for tectonic strain rates. So where is the BDT actually? For this we appeal to terrestrial seismology and the concept of the lithospheric strength envelope **(Figure 17)**. In the latter, the near surface (whether rock or ice) deforms by elastic-brittle failure (jointing and faulting), whereas at depth deformation is entirely due to viscous creep (by any number of mechanisms). At intermediate depths, however, there is a hybrid or mixed mode of deformation, combining ductile deformation and cataclasis, referred to as the semi-brittle regime (Kohlstedt et al., 1995; Kohlstedt & Mackwell, 2009; Scholz, 2019). The upper lithosphere is the seismogenic zone, to which faulting is usually restricted (e.g., Scholz, 1988; Scholz, 1998; Scholz & Contreras, 1998; Scholz, 2019; Molnar, 2020), and its lower boundary, at the transition to semi-brittle behavior, is taken as the true BDT (Kohlstedt & Mackwell, 2009); the transition to complete ductility (crystal plasticity) at even greater depth is referred to by Kohlstedt & Mackwell (2009) and Scholz (2019) as the (semi-)brittle-plastic transition (BPT).

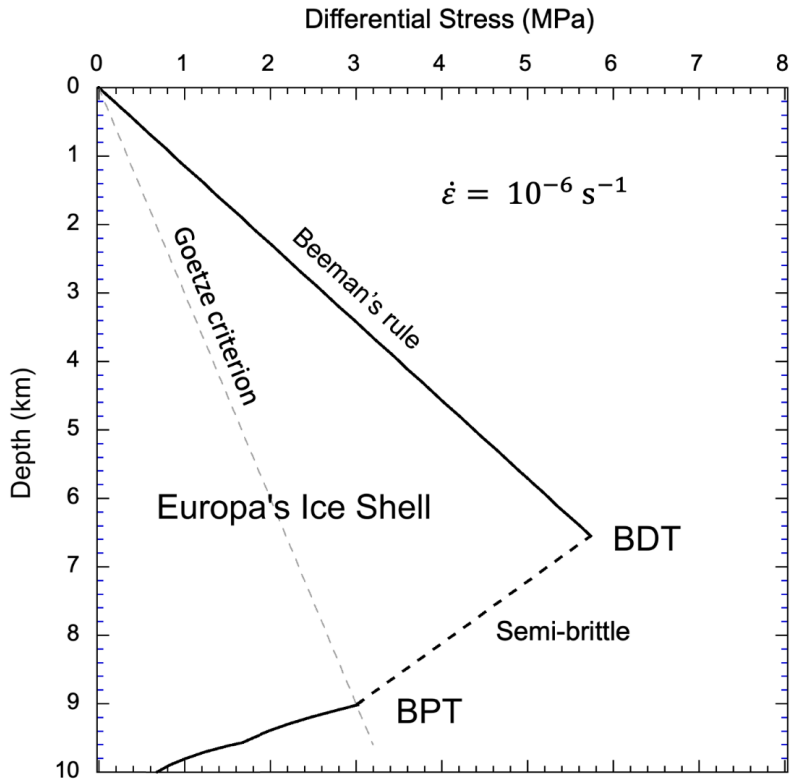

**Figure 17. Example strength envelope for Europa's ice shell in extension.** Here the shell is 10-km thick and conductive, with an exponential temperature profile from 100 K (surface) to 270 K (base). The strength of the upper, cold lithosphere is determined by the friction of pre-fractured ice (Beeman et al., 1988), whereas the brittle-ductile transition is initiated by dislocation creep in regime B (Durham & Stern, 2001). The transition to full plastic deformation (BPT) occurs close to where the differential stress for flow by higher-temperature dislocation creep regime A equals the confining pressure (the minimum compressive stress, in extension equal to the horizontal principal stress = $\sigma_{zz} - \Delta\sigma$). See Section 4.5 for additional details.

Our main point is that the relevant BDT in the present context is not the simple, high-stress "pointy" intersection of the brittle and ductile strength curves (Beeman et al., 1988), which has been used in all the past work interpreting impact ring graben, including our own (Singer et al., 2013b). Determining the transition to semi-brittle behavior for water ice, or indeed for any geologic material, is, unfortunately, non-trivial (Hansen, 2023). Although there are laboratory experiments on water ice that bear on this transition (Rist & Murrell, 1994; Golding et al., 2020), they are at much higher confining stresses than relevant here. So instead we follow the "rule of thumb" recommendation of Kohlstedt et al. (1995) and Kohlstedt & Mackwell (2009) that the BDT occurs at the depth and temperature where the differential stress required for ductile flow at a given strain rate is equal to the brittle yield stress (in our case, in extension), *divided by 5*. This is illustrated in **Figure 17** for an example conductive shell of 10 km thickness and relatively high strain rate of $10^{-6}$ s$^{-1}$.

With regard to the rule of thumb, the factor of 5 is empirical and qualitative, but motivated by numerous rock mechanics experiments and extensive field data. The BDT has been associated in crustal rocks with the onset of crystal plasticity in the weakest relevant major mineral (e.g., quartz in granite; Sec. 1.3 in (Scholz, 2019), but semi-brittle behavior is seen in any number of monomineralic rocks, e.g., halite salt-rock (Ding et al., 2021), quartz sandstone (Kanaya & Hirth, 2018), calcite limestone (Nicolas et al., 2017) and in water ice as well. Regarding hexagonal water ice I, it has a singular easy-slip direction, parallel to the basal plane (Glen & Perutz, 1954), so it is not hard to imagine that polycrystalline ice could deform, under appropriate conditions, by a mixture of crystal plasticity, brittle failure, and frictional sliding; see Petrenko & Whitworth (1999), their Sec. 8.3. The rule-of-thumb factor is expected to be composition dependent, but in the absence of an accepted, explicit constitutive model, we adopt the factor of 5 rule as a reasonable "guestimate."

Continuing, if the ring graben faults of Tyre and Callanish nucleated symmetrically at depth, the BDT is the logical limiting depth at which that could occur. If they originated at the surface, downward dynamic propagation could have continued beyond the BDT, into the semi-brittle zone, but not beyond the BPT. Here we only consider the BDT as this occurs at lower temperatures and thus provides a lower limit (floor) on Europa's heat flow at the time of impact. With our assumed dip angle above ($\theta = 62.3°$), the intersection depth $z = (w/2) \tan \theta$, where $w$ is the graben width, and the depth is thus equal to $0.95w$, or approximately the width of the graben. Using the average graben width from all of the measured locations at each impact site (**Table 1**), this minimum BDT depth ($z_{BDT}$) is estimated at $1.7 \pm 0.75$ km (1$\sigma$) for Tyre and $1.35 \pm 0.65$ km (1$\sigma$) for Callanish (**Table 2**). This is for the keystone geometry (**Figure 16a**). For the hourglass geometry (**Figure 16b**) $z_{BDT}$ might be up to twice as deep, or $3.4 \pm 1.5$ km and $2.7 \pm 1.3$ km for Tyre and Callanish, respectively. All these depth estimates are significantly lower than the shell thickness estimates described in Section 4.3, which has important implications for shell thermal structure, as discussed in Section 4.6 below.

### 4.5.1. Vertical Fractures at the Surface?

Here we briefly consider the possible role of vertical fractures at Europa's surface in accommodating the inferred normal fault offsets at Tyre and Callanish. It is not unusual for graben to be bounded, at the surface, by joint-controlled vertical faults. This is seen, for example, at Canyonlands, and at the East African and Icelandic rifts (e.g., Gudmundsson, 1992; McGill et al., 2000; Acocella et al., 2003; Kettermann et al., 2015). We are not suggesting that Europa's surface is rife with vertical joints such as might be formed in cooling basalt flows, but Europa's icy lithosphere is subject to tidal and tectonic cycles (Kattenhorn & Hurford, 2009), and even shock damaged but cold water ice may possess residual tensile strength under the dynamic conditions of ring graben formation. If, for example, we adopt the friction law determined by Beeman et al. (1988) for 77 K ice:

$$\tau = 0.55\sigma_n + 1.0 \text{ MPa} \quad , \tag{1}$$

where $\tau$ and $\sigma_n$ are the shear and normal stresses on a fracture surface, then Griffith failure theory would imply a tensile strength $\sigma_t \approx 0.5$ MPa. A surface undergoing horizontal extension would fracture perpendicular to the extensional stress direction and the fractures would only begin to rotate away from vertical at depths $h$ with sufficient overburden, or

$$h \approx 3\sigma_t/\rho g \quad , \tag{2}$$

where $\rho$ and $g$ are the ice density and surface gravity at Europa. For $\rho \approx 900$ kg m$^{-3}$ and $g = 1.314$ m s$^{-2}$, $h \approx 1.25$ km.

Conceivably, the Tyre and Callanish graben could be offset by such near-surface vertical tensile fractures (**Figure 16c**), or bounding faults that simply rotate to be closer to vertical at the surface, which would add directly to estimated depth to the BDT. This would nominally leave our estimates of the radial displacements and strain unchanged, however, at least for the idealized geometry shown. Initiation of normal faulting at Europa's surface by mode I (tensile) failure would also have the benefit of explaining 1) the circumferential nature of the graben (as opposed to spiral conjugate shear faults in plan due to compressive hoop, i.e., circumferential, stress; Allemand and Thomas, 1999) and 2) the great continuity along strike of many of the radially distant ones (at Tyre at least, whose whole geometry we can see).

The dynamic tensile strength of water ice at Tyre and Callanish is of course not well constrained. As noted earlier, the 1 MPa cohesion in Equation 1 may have been overly influenced by the mechanical stiffness of indium jacketing of the ice samples. In the shock physics calculations of Silber & Johnson (2017), they assumed a pre- and post-shock cohesion for water ice of 100 and 10 kPa, respectively, which would reduce the above offset to $z_{BDT}$ by 1 or 2 orders of magnitude, respectively. For simplicity, then, we will ignore this offset in Section 4.6 and those that follow.

### 4.5.2. Alternative Graben Origins?

Tectonic graben occur elsewhere in the Solar System and have been ascribed to a variety of causes. For the Moon, graben circumferential to impact basins are thought to have formed in response to mascon loading, due to flexure of the lunar lithosphere (Solomon & Head, 1980). To be clear, these graben formed well after the basin-forming impacts and are not themselves products of "ring tectonics," but the subsurface normal faulting mechanism or mechanisms are of interest here. Classically, the relatively uniform widths of many of these graben (termed "arcuate rilles" in older lunar literature) is explained by the bounding faults intersecting or originating at the base of porous and weaker impact-generated megaregolith (Golombek, 1979, 1985). Later papers argued, in contrast, that these graben were more likely generated by excess pressure at the tips of ascending but stalled-out basaltic magma dikes (Head & Wilson, 1993; Klimczak, 2014), taking a clue from dike swarms on the Earth and Venus (Ernst et al., 1995). This modest controversy continues (Martin & Watters, 2021), and it is instructive to consider whether either of these interpretations could apply at Tyre and Callanish.

In the latter case, if one imagines that all the europan ring graben are directly underlain by vertical magma-filled dikes, the negative buoyancy of the magma in question, namely water sourced from the ocean below, eliminates the excess pressure needed to drive graben formation. In the former case, there is no such impact-generated megaregolith of the required depth. One could appeal, perhaps, to up to several kilometers of tectonically-created fractures (Nimmo et al., 2003), but that would leave the question why the underlying solid lithosphere would undergo strains in excess of 1% during the impact modification stage and remain unfaulted. Plus, as one moves inward for either structure the disruption, displacement, and rotation of the crust becomes increasingly apparent, making a passive extension explanation for the ring-graben increasingly untenable.

Possibly, the two proposed mechanisms could be combined, in that through-going tension cracks could nucleate at the base of the ice shell and reach the base of a pre-fractured layer, whereby the extension of the separating ring segments then induce graben formation in the weaker surface layer above. Of course, why the tension fractures do not proceed all the way to the surface is unexplained, nor why they don't initiate at the surface to begin with, given that Europa's surface is replete with surface tension cracks/lineaments (Kattenhorn & Hurford, 2009). In addition, extending this concept to cover the much larger Valhalla and Asgard multiring basins or the furrow systems on Ganymede stretches credulity. Ultimately, we find little empirical support for these speculative graben-forming scenarios, but their saving virtue is that they are conceivably testable by future radar sounding data. Below we continue to interpret the europan ring-graben within the ring tectonics paradigm.

### 4.6. Strain Rates and Heat Flows

We visualize the ring-graben forming in sequence, from the inside out, via some combination of asthenospheric drag and gravity sliding. Below we solve for the surface heat flux ($q$) as a function of BDT depth for a variety of strain rates ($\dot{\varepsilon}$) potentially appropriate for post-impact graben formation. The transient crater collapse timescale can be used as an estimate for the shortest timescale relevant to graben formation (McKinnon & Melosh, 1980). For large gravity regime craters, the minimum collapse timescale is approximately equal to the period of a gravity wave with the same wavelength as the diameter of the transient crater, $\sqrt{D_{tr}/g}$ (McKinnon & Melosh, 1980; Melosh, 1982a). For Tyre and Callanish, with $D_{tr} \approx$ 20-to-23 km, this timescale is ≈125-to-130 s, or ~2 minutes. If this is the appropriate collapse timescale, then given the total strain estimates in **Figure 14**, which vary from 1–1.5% in the inner graben zone to ~0.1% at the periphery of both structures, then strain rates on the order of $10^{-4}$ s$^{-1}$ declining to ~$10^{-5}$ s$^{-1}$ at the periphery are appropriate. Vertical oscillation of the central cavity cannot be ruled out, as previously noted, so inward asthenospheric flow may not cease over a more extended period of time, by perhaps an order of magnitude or more. This would imply $\dot{\varepsilon}$ between ~$10^{-5}$ and $10^{-6}$ s$^{-1}$, or perhaps lower, in the graben formation zones of both Tyre and Callanish.

Alternatively, for another order-of-magnitude estimate, we note that strain rates decline with a $r^{-4}$ radial dependence for Newtonian inflow (Melosh & McKinnon, 1978; Allemand & Thomas, 1991), as long as the asthenospheric flow is not channelized into a horizontal layer. This assumption of a non-channelized asthenosphere is likely applicable to Europa given the presence of a deep (~150 km) ocean beneath "hot," subsolidus ice. Inverting the free collapse time above gives an $\dot{\varepsilon} \sim 10^{-2}$ s$^{-1}$ at the transient cavity rim. Tyre graben range from 45 to 95 km from the structure center (e.g., **Figures 6, 9b**), or 1.95 to 4.13 $D_{tr}$, implying potential graben formation strain rates of $3.3 \times 10^{-5}$ to $1.7 \times 10^{-6}$ s$^{-1}$, respectively. Callanish graben range from 30 to 75 km from the structure center, or 1.5 to 3.75 $D_{tr}$, implying potential graben formation strain rates of $1.0 \times 10^{-4}$ to $2.6 \times 10^{-6}$ s$^{-1}$, respectively. These strain rate ranges are close to those estimated in the previous paragraph, but may also be overestimates if the asthenospheric inflow takes longer to "settle down," as previously noted. In addition, Newtonian inflow is fine if the ocean dominates the inflow, but if most of the inflow occurs in warm, non-Newtonian ice (such as would apply to a thicker ice shell) then the radial dependence of the strain rate would be steeper and the strain rates implied for graben formation lower still. Accordingly, we will calculate temperatures at the BDT using $\dot{\varepsilon}$ that range between $10^{-4}$ and $10^{-8}$ s$^{-1}$, as shown on **Figure 18,** and for fiducial estimates below we use $10^{-6}$ s$^{-1}$ as a reasonable value that may have applied to much of the graben formation.

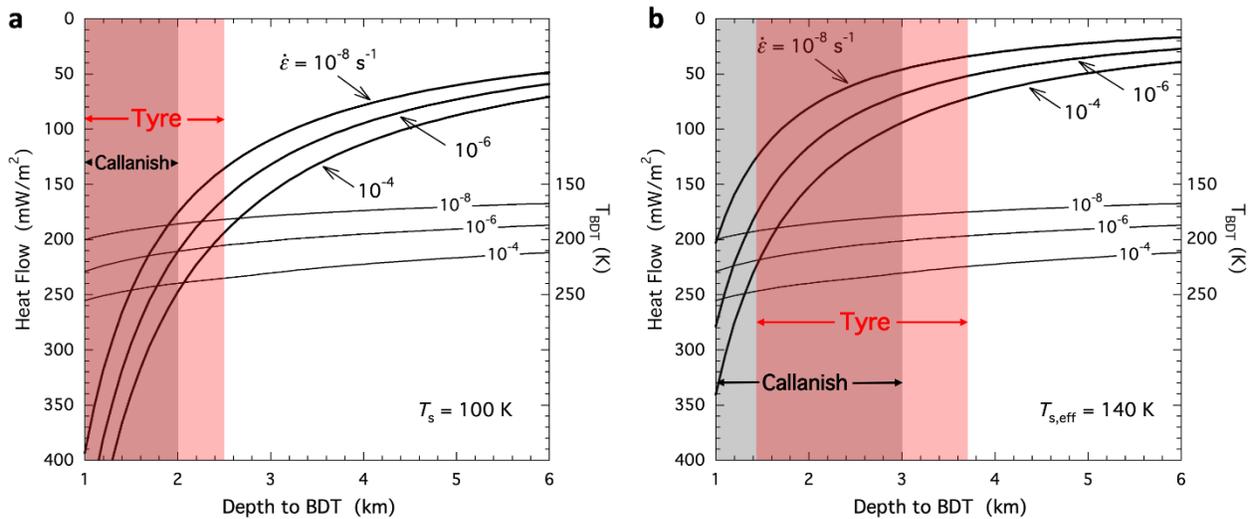

Figure 18. Heat flow from ring graben on Europa. (a) A range of heat flows are indicated based on *minimum* BDT depths and plausible asthenosphere strain rates (thick curves). Shaded regions indicate the range of possible minimum BDT depths for Callanish (gray) and Tyre (red). Values are calculated using the thermal conductivity of solid ice, which almost certainly overestimates the heat flows. Derived temperatures at the BDT are also shown (thin curves), which can be used

to calculate alternative heat flows. (b) Similar to panel a, but with BDT depths taken to be 1.5 × graben fault intersection depths (≈1.5$w$) and assuming an effective surface temperature of 140 K beneath a thin regolith cover. See text for details.

It is often said that the precise strain rate does not matter much in strength envelope calculations, but that is only true to a point. For example, Ruiz (2005), in an earlier but less detailed interpretation of Tyre graben, applied a geological strain rate of $10^{-14}$ s$^{-1}$. We would argue that this is too low to be relevant to the phenomena at hand. We further emphasize that we are assuming prompt, gravity-driven collapse of the transient cavities of Tyre and Callanish, such as seen in the shock physics calculations of Bray et al. (2014), Cox & Bauer (2015), and Silber & Johnson (2017). If, however, the asthenospheric ice in the lower portion of Europa's ice shell were sufficiently viscous (specifically, $\gtrsim 0.1 \rho g^{1/2} D_{tr}^{3/2} \sim 3 \times 10^8$ Pa-s), then the transient cavity would be overdamped and longer timescales and lower strain rates would apply (McKinnon & Melosh, 1980). We cannot disprove this possibility from observations alone (see Section 4.3), but consider it unlikely a priori (given the likely large driving stresses and strain rates for asthenospheric ice; see Fig. 2 of Durham & Stern [2001]).

As described earlier, we assume the BDT to, at minimum, represent the intersection of the brittle, frictional strength of ice and that of the relevant ductile flow law, subject to the "Kohlstedt criterion" above. From Beeman et al. (1988), their recommended cohesionless strength of fractured ice implies a differential failure strength in horizontal extension of

$$\Delta\sigma_{ex} = \frac{2\left[(\mu^2+1)^{1/2}+\mu\right]}{\{1+2\mu[(\mu^2+1)^{1/2}+\mu]\}}\rho g z = 0.724 \rho g z \qquad . \qquad (3)$$

Differential stresses due to ductile creep are given by

$$\Delta\sigma_{cr} = 0.2 \left(\frac{\dot\varepsilon}{A_m}\right)^{\frac{1}{n}} \left(\frac{2}{\sqrt{3}}\right)^{\frac{n+1}{n}} \exp\left\{\frac{Q}{nRT}\right\} \qquad , \qquad (4)$$

where the pre-factor $A_m$, stress exponent $n$, and activation energy $Q$ are experimentally determined, and the numerical factor $(2/\sqrt{3})^{(n+1)/n}$ relates the experimental geometry (cylinders under compression) to the plane strain conditions assumed here. For the differential stresses involved at Tyre and Callanish, dislocation mechanisms dominate. Values for $A_m$, $n$, and $Q$ can be found in Table 1 of Durham & Stern (2001): regime B for $T < 240$ K, regime A for 240 K $< T < 258$ K. High-temperature (pre-melting) creep for $T > 258$ K is also shown in **Figure 17** for completeness, but is not relevant to the BDT here.

For a given $\dot\varepsilon$ and $z = z_{BDT}$, equations 3 and 4 are set equal to determine the $T$ at the BDT. This in turn yields a heat flow estimate $q$ from

$$q = \frac{567 \text{ W m}^{-1}}{z_{BDT}} \ln\left(\frac{T_{BDT}}{T_s}\right) \qquad , \qquad (5)$$

where $T_{BDT}$ and $T_s$ are the temperatures at $z_{BDT}$ and the surface, respectively. We realize that equation 5 incorporates the oft-quoted thermal conductivity ($k$) for solid water ice of 567 W m$^{-1}$/$T$ from Klinger (1980), and thus likely leads to an overestimate of heat flows. For an airless body such as Europa, unclosed fractures in the upper lithosphere (e.g., Nimmo et al., 2003; Nimmo & Manga, 2009) must reduce conductive heat transport, as cold fracture surfaces are vacuum interfaces across which heat is only conducted at asperities. We take this into consideration in what follows as well.

**Figure 18a** shows a strong dependence of derived heat flow as a function of $z_{BDT}$. For the range of $z_{BDT}$ values quoted above (also see **Table 2**) the heat flows $q$ are all high, >100 mW m$^{-2}$, and for the average $z_{BDT}$ (or graben width $w$) values, extraordinarily so, ≳200 mW m$^{-2}$ for both Tyre and Callanish. Are such high values plausible? We would argue not. First, they would imply either a thin conductive ice shell or stagnant lid only ~3 km thick (next section), which is ruled out by the impact simulations to date (described earlier). Specifically, we regard impact simulations as the most direct, relevant constraint on Europa's ice shell thickness with time, as the physics of shock wave propagation and the excavation stage are particularly well understood. The ensemble of europan crater depths in Schenk (2002), which for $D$ < 15 km were not selected for relative youth but rather by the availability of appropriate images for shadow-length measurements, is simply incompatible with a 3-km-thick floating ice shell.

Furthermore, it should be obvious that the entire scatter of graben widths as a function of radial distance in **Figure 9** *cannot* represent actual spatial variations in $z_{BDT}$, though such may contribute. The variation in graben width is strong, up to a factor of five for Tyre, in some instances along strike of a single graben. At a given time in Europa's history, the long-wavelength variation in Europa's shell thickness should be, nominally, modest (Nimmo et al., 2007), but features such as chaos or the oft-hypothesized intercrustal water sills could provide variations in heat flow on much shorter length scales (Collins & Nimmo, 2009). But such features are by their very nature geologically short-lived and it is implausible that there would be so many extant at any one time at Tyre and Callanish sites. Nor is there any geological evidence at Tyre in particular that this was the case, in those areas of graben formation not blanketed by ejecta. (Tyre appears to have formed in typical background ridged plains [Kadel et al., 2000]). There is the possible influence of pre-existing crustal faults. But this mostly begs the question of the explicit source of graben width variation, and how best to interpret the data at hand with respect to heat flow (i.e., should one focus on a data set average, average width as a function of radial range, maximum widths, or something different?).

Even for constant or slowly varying $z_{BDT}$ with position, geological variability may arise from, for example, details of the nucleation of the secondary or antithetic bounding fault in the hourglass model (**Figure 16b**), because fault intersections need not meet at the BDT, or from azimuthal variation in strain rate. In our view, the lower $z_{BDT}$ values, at a given $r$, almost certainly represent fault intersections above the BDT (the X configuration). Mean or modal $z_{BDT}$ values are probably a better measure, but even these are best regarded as minimum values.

Accordingly, we also list in **Table 2** average $z_{BDT}$ values from 1.5 × the graben bounding fault intersection depths, which we simply label as "plausible" BDT depths.

For average Tyre BDT depths of, say, 3 km (which would appear quite appropriate in the inner graben zone; see **Figure 9b**), a surface temperature of 100 K, and solid ice conductivity, the implied heat flows are ~135 ± 25 mW m$^{-2}$ for the range of $\dot{\varepsilon}$ in **Figure 18a** (cf. **Table 2**, which lists $q$ for a plausible BDT depth of 2.6 km). But even this value is likely an overestimate because the thermal conductivity is a maximal (solid ice) value. If the thermal conductivity were half that of solid water ice, then this heat flow estimate is halved. As far back as the beginning of the *Voyager* era, Shoemaker et al. (1982) argued that the effective surface temperature of Ganymede was as high as 130 K, a temperature at which both sintering and water vapor transport in that body's regolith provided sufficient thermal communication to justify using solid conductivity values in models. That is, temperatures should increase very rapidly with depth in a low-conductivity regolith until $T(z)$ ~130 K is reached. A detailed calculation of viscous pore (and crack) closure due to overburden pressure, for water ice on Europa, by Nimmo et al. (2003) gave $T$ ~140 K as the limiting temperature above which ice should act to remain solid and thus fully conductive on a geologic timescale. Following Shoemaker's argument, but using 140 K as the effective $T_{s,eff}$, gives a heat flow from Equation 5 of ~70 ± 25 mW m$^{-2}$ for $T_{BDT}$ appropriate to $z_{BDT}$ = 3 km (**Figure 18b, cf. Table 2**). (Note that $T_{BDT}$ values are also plotted in **Figure 18** for the same range of strain rates; the reader can apply their own ice conductivity model if they so wish).

For Callanish, BDT depths are generally smaller and derived heat flows higher than at Tyre. For our plausible $z_{BDT} \approx 2$ km (also quite appropriate to the inner graben zone of that structure, given the range of graben widths there; see **Figure 9b**), the implied heat flows are ~210 ± 35 mW m$^{-2}$ (**Figure 18a**, **Table 2**) for full solid ice thermal conductivity, or ~115 ± 35 mW m$^{-2}$ if the effective $T_{s,eff}$ = 140 K (**Figure 18b**). We discuss the similarities and differences between Callanish and Tyre in greater detail in Section 5.

To all this we can add the possibility that the brittle portion of the ice lithosphere retains some cohesion post-impact (perhaps similar to that in Equation 1). In this case the brittle strength in **Figure 17** moves to the right, which for the same $z_{BDT}$ and $\dot{\varepsilon}$ implies a *lower* $T_{BDT}$. At the same time, this increases the likelihood of steeper bounding faults at the surface and larger $z_{BDT}$ values. Both physical effects imply still lower heat flow values than we have just presented.

In summary then, the range of observed graben widths at the Tyre and Callanish impact structures (**Figure 9**) imply a correspondingly shallow range of $z_{BDT}$ values. These in turn imply remarkably high heat flow values at the times of their respective impacts, in the simplest interpretation. Various physical effects (subsurface fault and thermal conductivity structure) likely reduce the true $q$ values to ~100 mW m$^{-2}$ or less, however. These lower $q$ values are compatible with other geophysical estimates from folding or (some) flexure features on Europa (Dombard & McKinnon, 2006; Nimmo & Manga, 2009). Unfortunately, we cannot be more accurate because of the considerable uncertainties involved, in structural interpretation, brittle yield strength, and appropriate ice thermal conductivity, as well as incomplete *Galileo* coverage at Callanish. There are also formal uncertainties in measured flow parameters for the ductile creep laws (Durham & Stern, 2001), but these are less important than the larger, systematic uncertainties just discussed. We also note that all the above analyses do not depend on precise values of the Tyre and Callanish transient craters, nor on the water-ice grain size, though obviously if the composition of the shell were substantially different from water ice, so would be the implications.

**Table 2. Derived Quantities**

|  | **Tyre** | **Callanish** |
|---|---|---|
| Maximum radial strain | 1–1.5% | 0.6–1.2% |
| Minimum BDT depth ($z_{BDT}$) (from avg. width & 1 × fault intersection depth) | 1.7 ± 0.75 km | 1.35 ± 0.65 km |
| Plausible BDT depth ($z_{BDT}$) (from avg. width & 1.5 × fault intersection depth) | 2.6 ± 1.15 km | 2.0 ± 1.0 km |
| Heat flow ($q$)[a] Minimum total shell thickness ($z_{shell}$)[a] (for solid-ice $k$, $T_s$ = 100 K & plausible BDT depth) | ~160 ± 30 mW m$^{-2}$ ~3.7 ± 0.7 km | ~210 ± 35 mW m$^{-2}$ ~2.7 ± 0.5 km |
| Heat flow ($q$)[a] Minimum total shell thickness ($z_{shell}$)[a] (for solid-ice $k$, $T_{s,eff}$ = 140 K & plausible BDT depth) | ~85 ± 30 mW m$^{-2}$ ~4.8 $^{+2.0}/_{-1.5}$ km | ~115 ± 35 mW m$^{-2}$ ~3.3 $^{+1.4}/_{-0.8}$ km |

[a] Uncertainties based on assumed strain rates during graben formation of $10^{-8}$ to $10^{-4}$ s$^{-1}$

### 4.7. Full Ice Shell Thickness and Ring Tectonics

The BDT depths represent the brittle-elastic shell thickness, but full shell thicknesses ($z_{shell}$) can also be estimated. At the high strain rates associated with timescales of the transient crater collapse, the temperatures at the BDT are relatively high. Simply extrapolating the $T_{BDT}$ values in **Figure 18a** to the depth of melting (at 273 K), assuming a conductive as opposed to a convective temperature profile, and solid ice conductivity, results in estimates of 4.3 ± 0.8 and 2.7 ± 0.5 km total shell thicknesses, for Tyre ($z_{BDT}$ = 3 km) and Callanish ($z_{BDT}$ = 2 km) respectively (cf. **Table 2**), with the range in $z_{shell}$ dependent on $\dot{\varepsilon}$. Alternatively, adopting an effective $T_{s,eff}$ = 140 K (**Figure 18b**) gives $z_{shell}$ = 5.5 $^{+2.7}/_{-1.5}$ and 3.3 $^{+1.4}/_{-0.8}$ km for Tyre ($z_{BDT}$ = 3 km) and Callanish ($z_{BDT}$ = 2 km), respectively. If the ring graben formed over longer, more geological time scales, thicker shells would be indicated as well, but we don't advocate such here.

These shell thickness estimates, even for Tyre, are barely consistent with the constraints on shell thickness derived by numerical impact calculations of ~7-to-10 km (Bray et al., 2014; Cox & Bauer, 2015; Silber & Johnson, 2017). Perhaps the wider graben seen at both structures (≳4 km at Tyre; **Figure 9**) are a better indicator of absolute shell thickness. We caution, however, that the wider troughs may be more complicated structurally, i.e., rift-like as opposed to simple graben, but we do not have the image resolution or other data to be definitive. Our shell thickness estimates are, however, closer to a ~5-to-7 km thick conductive layer, or stagnant lid, atop a thicker convecting ice shell, which also matched the Schenk (2002) depth/diameter data in the impact simulations of Silber & Johnson (2017). Thickness estimates for a conductive lid overlying 260 K ice (McKinnon, 1999) are only reduced ~10% from the above $z_{shell}$.

Regardless, using a depth-to-diameter assumption of 0.5 for the transient crater cavity, the transient crater depths would have been ~10-to-11.5 km for Callanish and Tyre, respectively. This implies the formation of either structure should have completely pierced the ice shell or stagnant lid at the time, depending on whether the ice shell was conductive throughout or convecting at depth. This satisfies an important constraint of the ring tectonics model (McKinnon & Melosh, 1980).

Another requirement that must be met is sufficiently large basal stress from the asthenospheric inflow. In the thin elastic-plastic lithosphere model of (Melosh, 1982b), the character and extent of any circumferential graben zone is controlled by a dimensionless parameter

$$\gamma = \frac{\sqrt{3} S D_{tr}}{4 Y z_{lith}}, \qquad (6)$$

where $S$ is a measure of the basal traction, $D_{tr}$ is the transient cavity or crater diameter, and $Y$ and $z_{lith}$ are the yield strength and thickness of the ice lithosphere. In Melosh (1982b), $\gamma$ should be $\gtrsim 2$ for failure and nominally larger ($\gtrsim 5$) for graben to form. This analytical model supposes that the failed lithosphere on a broad scale can be represented as a continuum at its plastic failure limit, rather than a brittle-elastic-ductile layer with a depth-dependent failure stress $\Delta\sigma(z)$ (**Figure 17**). As such, the product $Yz_{lith}$ is roughly equivalent to $\int_0^{z_{lith}} \Delta\sigma \, dz \sim 0.5(\Delta\sigma)_{BDT} z_{BDT}$. For $S \sim 0.1 \rho g H$, where $H$ is the transient cavity depth (Melosh, 1982b), equal in our case to $D_{tr}/2$ ~10 km, $\gamma \sim 1.7 \times 1.3$ MPa $\times$ 10 km/(4 $\times$ 1.0 MPa $\times$ 2.5 km) = 2.2, where we have used 2.5 km as a $z_{BDT}$ representative of both Tyre and Callanish. Thus, these impact structures appear marginally capable of exhibiting ring graben, to the degree that the plastic lithospheric failure theory of Melosh (1982b) can capture the actual behavior of a brittle-elastic-ductile lithosphere that forms normal faults at discreet spacing. The scaling of $\gamma \sim (D_{tr}/z_{BDT})^2$ is, however, notable in that it is a strong function of impact size, which may explain the abrupt appearance of impact rings with increasing crater scale on Europa. And if we have underestimated $D_{tr}$ by 50%, then the specific Melosh criterion for an annular ring of graben would be met.

Ring graben spacing also deserves comment. Although we do not emphasize it in this work, the general increase in inter-graben spacing with radial distance (seen best at Tyre; **Figure 11b**) is consistent with the ring tectonics model. We expect driving stresses to decline with radial distance, whether from asthenospheric drag or from extensional, radial membrane stress transmitted from the lithosphere closer to the collapsing cavity. The stress release as each graben forms, more or less in sequence from the inside out, should be effective over a radial distance equal to at least several graben bounding fault intersection depths (≈observed widths) (e.g., Melosh & Williams, 1989; cf. Lachenbruch, 1961). In other words, we do not expect the ring graben to form on top of one another, and the spacing of the innermost grabens may also be controlled by the depth to the BDT.

## 5. Tyre and Callanish Compared

In addition to being located on different hemispheres of Europa (their centers are ~172° of longitude apart) and their slightly different diameters, Tyre and Callanish also formed at different times. Several prominent double ridges and fractures have been previously noted to superpose Tyre, which itself appears to have formed on ridged plains (e.g., Kadel et al., 2000; Moore et al., 2001; Bierhaus et al., 2009). These are especially prominent in the lower phase angle (high sun) *Galileo* images (Moore et al., 1998) as shown in **Figure 4a**. For the part of Callanish that was well imaged, there are no obvious interruptions of the central chaotic zone or inner graben by later ridges or fractures, but there are a few possible superpositions on the very outer graben. Given the more pristine nature of Callanish, it is likely somewhat younger than Tyre. At least one of Callanish's rings superposes a small chaos region, and chaos formation is nominally typical of Europa's later evolution (Collins & Nimmo, 2009). Callanish also generally appears darker in low resolution higher sun images than Tyre. Overall the surface of Europa is quite young, estimated to be an average of 20-200 Ma (Bierhaus et al., 2009; cf. Nesvorný et al., 2023) based on global crater data. Tyre and Callanish can't be dated precisely but their formations could be separated by 10s or even a hundred million years (if Tyre formed in the first half of Europa's recorded geologic history, and Callanish in the second).

The differences in ring-graben widths between the two structures are notable. The Callanish trough/graben are narrower as an ensemble (though there are overlaps in the distributions; **Figure 9**), so the inferred $z_{BDT}$ are shallower and the derived heat flows at the time of impact higher for Callanish. This is opposite to the "received wisdom" regarding Europa's geological history, in which Europa's shell is held to have thickened over time due to declining heat flow, promoting an extensional tectonic environment overall, and resulting in eras dominated progressively by the formation of ridged plains, then bands, and ultimately chaos (e.g., Greeley et al., 2004; Kattenhorn & Hurford, 2009; McKinnon et al., 2009; Nimmo & Manga, 2009). On the other hand, tidally heated convecting ice shells can evolve in non-intuitive ways, so simultaneous shell thickening and increased heat flow is possible (Barr & Showman, 2009), and the different crater-forming responses at Callanish and Tyre may be key evidence in this regard.

The two impact structures are similar enough in scale (the transient diameters are within 10-15% of each other), so a systematic variation in strain rate is unlikely as an explanation for the difference in the heat flows we derive (and works in the wrong direction anyway: the collapse timescale would be shorter for a smaller impact, thus the strain rates higher near the impact, and for the same heat flow that would produce a deeper [not shallower] depth to the BDT [see **Figure 18**] for Callanish). We do note that a decrease in graben width with radial distance (as a running average in **Figure 9b**) for both structures is qualitatively consistent with the argument first made by Allemand & Thomas (1991) for decreasing strain rates with radial distance. The variation in average graben width with distance is, however, stronger than predicted by the variation in strain rates estimated above for the same heat flow, even for larger $z_{BDT}$ (right hand side of **Figure 18**). This could imply a steeper radial decline in strain rate than estimated, one more consistent with non-Newtonian inflow of warm, ductile ice during transient cavity collapse.

We also considered the hypothesis that spatial variation in heat flows, rather than temporal variation, could be an alternative explanation for the inferred higher heat flow at Callanish. Callanish is somewhat closer to the equator (-16.7º S, 334.5º W) than Tyre (+33.6º N, 146.6º W), but the implied difference in subsurface average temperature is only a few K at best. Spatial variations in tidal heating may be pronounced, however (Ojakangas & Stevenson, 1989), even if lateral flow of warm ice at depth acts to even out differences in shell thickness (Nimmo et al., 2007). Whether pronounced variations in surface heat flow result is less clear: for example, Tobie et al. (2003) suggest not for a convecting icy shell. In any event the present positions of Tyre and Callanish do not suggest strongly divergent heat flows, based on tidal flexing patterns (e.g., Ojakangas & Stevenson, 1989). If spatial variation was the cause of the differences between Tyre and Callanish, that would likely require, at minimum, that these impacts formed at suitably different latitudes and longitudes than today and that the ice shell has since reoriented (i.e., by non-synchronous rotation or true polar wander).

## 6. Conclusions

The collapse of large transient impact cavities may lead to the creation of one or more exterior rings. The existence and extent of such ring systems depend on the thickness of the mechanical lithosphere at the time and place of impact. Icy satellites offer a valuable laboratory to explore this paradigm. The two largest impact structures on Europa are Tyre (≈190 km across, when measured to its outermost rings) and Callanish (≈130 km across, though incompletely seen). Tyre and Callanish possess compact systems of circumferential graben-like troughs, essentially miniature versions of the much larger Valhalla and Asgard/Utgard multiringed structures on Callisto, and plausibly related to the hemisphere-scale furrow systems on Ganymede. The compact nature of the europan ring structures implies a relatively thin icy lithosphere, consistent with steep temperature gradients due to strong tidal heating at the time and locale of impact. Measurements of trough depths allow us to determine the inward radial strain represented by the graben sets. These values are high enough (>1%) that, along with observations of inward displacement of discrete, arcuate ice shell fragments (by several kilometers), decoupling or ductile deformation at depth is strongly implied.

We use measurements of trough width and depth to constrain heat flows at the time of impact. Assuming the troughs are graben, we find that fault intersection depths are equal to graben width (to within 10%). Assuming that the bounding faults originated or nucleated at the brittle-ductile transition (BDT) allows us in principle to relate this depth to the surface heat flux through the temperature profile in the ice shell and the rheology of water ice. The BDT occurs at a depth and temperature where the differential stress required for ductile flow at a given strain rate is a specified fraction of the brittle yield stress. For the high strain rates appropriate to post-impact crater collapse, we find anomalously high heat flows, $\gtrsim 200$ mW m$^{-2}$. We argue, however, on the basis of terrestrial field studies and numerical models, that the "hourglass" model of graben formation is likely more correct than the classic "keystone" model. If so, our initial BDT depths are likely underestimates. Furthermore, preexisting fractures imply lower thermal conductivity for cold lithospheric ice, and the combination implies true or at least more plausible heat flows ≲100 mW m$^{-2}$, similar to many other (though not all) literature estimates for Europa.

BDT depth estimates are shallower and heat flow estimates are higher for Callanish, the younger of the two structures, which is contrary what is usually assumed for the directionality of Europa's recorded geologic history. Extrapolating from the range of plausible heat flows for both structures we estimate total conductive ice shell thicknesses—or stagnant lid thicknesses, if the ice shell is convecting at depth—roughly between 2.5 and 8 km. Such thicknesses are compatible (at the high end) with previous numerical calculations of impact morphology and morphometry on Europa. Unfortunately, fundamental uncertainties in graben subsurface fault geometry, the appropriate brittle yield strength for Europa's water ice, and ice thermal conductivity as a function of depth preclude a more definitive evaluation. Qualitatively, the appearance of the two multiring "basins" is consistent with impact breaching of Europa's ice shell, but quantitatively, we slightly favor an interpretation of a conductive stagnant lid over deeper convecting ice. At our level of analysis, the ring graben patterns and topography do not clearly discriminate between conductive and convective shells.

We do point out that all of the published numerical impact simulations to date (Bray et al., 2014; Cox & Bauer, 2015; Silber & Johnson, 2017) also advocate either thinner conductive shells or higher heat flows than what are, for the most part, considered canonical for Europa (cf. Billings & Kattenhorn, 2005; Nimmo & Manga, 2009; Howell, 2021). One can always appeal to specific formation circumstances on Europa to explain locally thin lithospheres and high heat flows (i.e., at double ridges), but impacts, including Tyre and Callanish, are (mostly) agnostic as to where they form, so this discrepancy is a bit of a puzzle, and worthy of deeper examination.

Further progress, in the near term, will likely come from additional numerical simulations, particularly those targeted at Tyre and Callanish (e.g., Silber & Johnson, 2018). Such would be especially valuable if carried out with enough fidelity to resolve such structural features as the bounding faults of ring graben. Over the longer term, resolution of fundamental issues will require new surface and subsurface data from Europa. NASA's *Europa Clipper* mission is slated to overfly both Tyre and Callanish in the early 2030s. Along with complete and higher-resolution image coverage and topographic and compositional mapping, ice-penetrating radar sounding has the potential to definitively prove or disprove the hypotheses advanced here (see, e.g., Sbalchiero et al., 2023). Ice penetrating radar has the possibility to provide a much clearer picture of the subsurface structures in the realm of the ring graben and bear on whether the graben fractures may facilitate material communication between the surface and the ocean.

**Acknowledgments**

This work was originally supported by NASA Planetary Geology and Geophysics grant NNX11AP16G (to WBM), NASA Earth and Space Science Fellowship NESSF (to KNS), and the Lunar and Planetary Institute (PMS). Sincere thanks go Veronica Bray and an anonymous reviewer for their detailed comments, which resulted in an improved and much more complete presentation. We thank Kimberly Lichtenberg and Amy Barr for initial work on this project. We also thank Mike Bland for assistance with projected *Galileo* and *Voyager* images of Europa, C. Koeberl for bringing the Praia Grande feature to our attention, and most of all the late H. J. Melosh for his keen insights into all things impact. We look forward to *Europa Clipper* data on the surface and subsurface structure of these two remarkable impacts.

**Open Research**

All graben measurements and data displayed in the above plots are available at a figshare.com repository under this DOI: 10.6084/m9.figshare.22336048 under at CC BY 4.0 license (Singer et al., 2023). The data is available as excel files, .csv files, and the mapped graben outlines and center points are available as ArcGIS shapefiles. The raw *Galileo* and *Voyager* images of Europa are available at the Planetary Data Systems Cartography and Imaging node https://pds-imaging.jpl.nasa.gov/. GIS-ready *Galileo* images and mosaics tied to a common basemap are available in Bland (2021).

Singer, McKinnon, and Schenk, 2023 – Europa Impact Structure Ring-graben